\documentclass[letterpaper,onecolumn,english,pre]{revtex4}

\usepackage{amsmath,amssymb,amsthm}

\usepackage{hyperref}
\usepackage{amsthm}
\usepackage{graphicx}
\usepackage{subfigure}
\usepackage{color,bbm}
\usepackage[normalem]{ulem}

\usepackage{amsfonts}
\usepackage{mathrsfs}

\usepackage{color}

\usepackage{amsthm}

\begin{document}
\date{\today}
%\title{\bf  How finite diffusion affects the current of the totally asymmetric simple exclusion process }
\title{\bf  Modelling the effect of ribosome mobility on the rate of protein synthesis  }
%\title{\bf  Modelling active transport of molecular motors coupled with diffusion in confined media }

\author{{\normalsize{ Olivier Dauloudet${}^{1,2,\dagger}$, Izaak Neri${}^{3,\dagger,\ast}$, Jean-Charles Walter${}^{1,\dagger,\ast}$,  J\'{e}r\^{o}me Dorignac${}^{1}$, Fr\'{e}d\'{e}ric Geniet${}^{1}$,  Andrea Parmeggiani${}^{1,2}$}}}
\email{izaak.neri@kcl.ac.uk}
\email{jean-charles.walter@umontpellier.fr}
\email{andrea.parmeggiani@umontpellier.fr}
\affiliation{\noindent    ${}^1$\textit{ Laboratoire Charles Coulomb (L2C), Montpellier University, CNRS, Montpellier, France}}
\affiliation{\noindent  ${}^2$\textit{Laboratory of Parasite Host Interactions (LPHI), Montpellier University, CNRS, Montpellier, France}}
\affiliation{\noindent  ${}^3$\textit{ Department of Mathematics, King’s College London, Strand, London, WC2R 2LS, UK}}
\affiliation{\noindent   ${}^\dagger$These authors contributed equally.}

\begin{abstract}      
Translation is one of the main steps in the synthesis of proteins.  It consists of ribosomes that translate    sequences of nucleotides encoded on  mRNA into  polypeptide sequences of amino acids. Ribosomes bound to mRNA move unidirectionally, while unbound ribosomes  diffuse  in the  cytoplasm.   It  has been hypothesized  that finite diffusion of ribosomes plays an important role in ribosome recycling and that  mRNA circularization enhances the efficiency of translation, see e.g.~Ref.~\cite{LodishEight}. In order to estimate the effect of cytoplasmic diffusion on the rate of translation, we consider  a 
Totally Asymmetric Simple Exclusion Process (TASEP) coupled to a finite diffusive reservoir, which we call the 
Ribosome Transport model with Diffusion (RTD).   In this model, we derive an analytical expression for the rate of protein synthesis  as a function of the diffusion constant of ribosomes, which is corroborated with results from  continuous-time Monte Carlo simulations. Using a wide range of biological relevant parameters, we conclude that   diffusion is  not a rate limiting factor in translation initiation because diffusion is fast enough in biological cells.
\end{abstract}

%Translation is one of the main steps in the synthesis of proteins. It consists of ribosomes that translate sequences of nucleotides encoded on  mRNA into polypeptide sequences of amino acids. Ribosomes bound to mRNA move unidirectionally, while unbound ribosomes diffuse in the  cytoplasm. It has been hypothesized  that finite diffusion of ribosomes plays an important role in ribosome recycling and that mRNA circularization enhances the efficiency of translation. In order to estimate the effect of cytoplasmic diffusion on the rate of translation, we consider a Totally Asymmetric Simple Exclusion Process (TASEP) coupled to a finite diffusive reservoir, which we call the Ribosome Transport model with Diffusion (RTD). In this model, we derive an analytical expression for the rate of protein synthesis as a function of the diffusion constant of ribosomes, which is corroborated with results from continuous-time Monte Carlo simulations. Using a wide range of biological relevant parameters, we conclude that diffusion in biological cells is fast enough so that it does not play a role in controlling the rate of translation initiation.

\maketitle

%--------------------------------------------------------------------------------------------------
\section{Introduction}       
%--------------------------------------------------------------------------------------------------
Cells synthesize proteins by first transcribing the hereditary information encoded in genes into functional mRNA, and subsequently by translating the mRNA nucleotide sequence into polypeptide sequences~\cite{LodishEight}. The translation of mRNA into a polypeptide sequence can be divided into three stages, namely, the initiation, elongation and termination stages \cite{LodishEight}.   During initiation,  a ribosomal complex (consisting of two ribosomal subunits, initiation factors, and tRNA)  is assembled at the 5' end of a mRNA chain. After initiation, the ribosomal complex moves (or elongates) from the 5' end towards the 3' end of the mRNA while forming a polypeptide chain. In the final termination stage, the ribosome complex releases the polypeptide chain,   unbinds from the mRNA and dissasembles.

Translation is mainly controlled at the initiation step, as it is the rate limiting step in translation~\cite{Kozak,Pestova,Hershey,Shah}. Initiation is a complex process involving several molecular actors, and  it is therefore difficult to understand  all the molecular mechanisms that are relevant for translation control.  Nevertheless, coarse-grained mathematical modelling  can uncover  which physical mechanisms play a role in translation control.    

It has been argued that the recycling of ribosomes through Brownian diffusion in the cytosol plays an important role in the control or regulation of translation~\cite{LodishEight, Chou2003, Fer2017, Lucas2019}. When a ribosome unbinds from the mRNA after termination, it can either rebind to the same mRNA or bind to another mRNA. If the diffusion of ribosomes is slow enough, then circularization of the mRNA  could enhance the rate of ribosome recycling through cytosolic diffusion~\cite{LodishEight, Chou2003, Wells, Vicens}. On the other hand, this effect would be negligible if diffusion of ribosomes is fast enough. In this paper we use physical modelling to determine whether recycling of ribosomes through diffusion can play a role in controlling mRNA translation.

In order to study how ribosome mobility affects the mRNA initiation rate and thus the protein production, we present a minimalistic physical model that describes both the translation of mRNA by ribosomes and  the diffusion of ribosomes in the cytoplasm. We call this model the Ribosome Transport model with Diffusion (RTD). From a physical viewpoint, the RTD consists of particles (the ribosomes) that diffuse in a box and can bind to a one-dimensional substrate (mRNA). Particles bound to the substrate move unidirectionally and cannot overtake. The RTD consists thus in a Totally Asymmetric Simple Exclusion Process (TASEP)~\cite{Blythe} in contact with a diffusive reservoir. If diffusion is fast enough, then we recover the standard TASEP model, which describes in detail the elongation stage of mRNA translation~\cite{Macdonald1968, Macdonald1969, Rom2009, Bonnin2017}. On the other hand, when diffusion is slow, then a concentration gradient is formed in the reservoir and there will be a tight coupling between active transport on the filament and diffusion in the reservoir.
In this regime, the RTD  describes the interplay of active and passive transport in cellular media, leading  to the formation of a gradient of molecular species.   Phenomena of active transport coupled to a diffusive reservoir have been  studied before in the literature, see for example Refs.~\cite{Chou2003, Klumpp2, Klumpp1, Muller2005, Neri2, Neri1, Cian, Graf, Rank, Fer2017, Lucas2019, Verma, Jindal,Dauloudet,Parmeggiani09}. In these studies, much focus has been put  on nonequilibrium phase transitions~\cite{Blythe, Chou2011, Chow2013, Neri1}.

In the present paper, we use mean-field theory to derive an analytical expression  for the protein synthesis in the  RTD model, which is corroborated with numerical results obtained from  continuous-time Monte Carlo simulations. Subsequently,  we use the analytical  expression  for the protein synthesis rate to  discuss the biological relevance of Brownian diffusion in ribosomal recycling.   By considering a broad range of biological parameters, we come to the conclusion that under physiological conditions finite diffusion of ribosomes is not important in the control of mRNA translation. Thus, circularisation should not occur in order to prevent the limiting effect of Brownian diffusion of ribosomes in the cytoplasm on initiation of translation~\cite{LodishEight, Chou2003, Wells, Vicens}.     In addition, we discuss how the spatial dimensions of the reservoir and geometry impact the protein synthesis rate and we find  qualitative difference in the dependence of the   protein synthesis rate on the length of the mRNA between two and three dimensions. Both cases are biologically relevant: the three-dimensional case applies to cytoplasmic translation, whereas the two-dimensional case applies to endoplasmic reticulum translation. 
%e come to different conclusions than Refs.~\cite{Fer2017, Lucas2019}: according to our analysis of the RTD model, 
%We find a qualitative difference between two and three dimensions in the dependency of the protein synthesis rate on the length of the mRNA. We also discuss the effect of a finite volume on the protein synthesis rate.  %In addition, in the present paper we discuss  mean-field theory in both two and three dimensions  and we compare theoretical results with  simulation results.   
% ??? put a comment on the saturation of J vs D: in Fig.4, why don't we get any saturation ?    

The paper is organized as follows. In Sec.~\ref{Sec:def}, we define the RTD model.   In  Sec.~\ref{Sec:theory}, we present a mean-field theory for the RTD model and derive analytical expressions for the   protein synthesis rate as a function of the diffusion coefficient of ribosomes. In Sec.~\ref{Sec:Simulation}, we compare theory with simulations results using a continuous-time algorithm. In Sec.~\ref{Sec:param}, we discuss the biological relevance of the model.  We conclude the paper with a discussion in  Sec.~\ref{Sec:disc}, and in Appendix~\ref{appendixA} we present analytical results for the  concentration profile of ribosomes in the cytoplasm.  

%--------------------------------------------------------------------------------------------------
\section{Model definition: Ribosome Transport with Diffusion} \label{Sec:def}  
%--------------------------------------------------------------------------------------------------

\begin{figure}
{\includegraphics[width=0.5\textwidth]{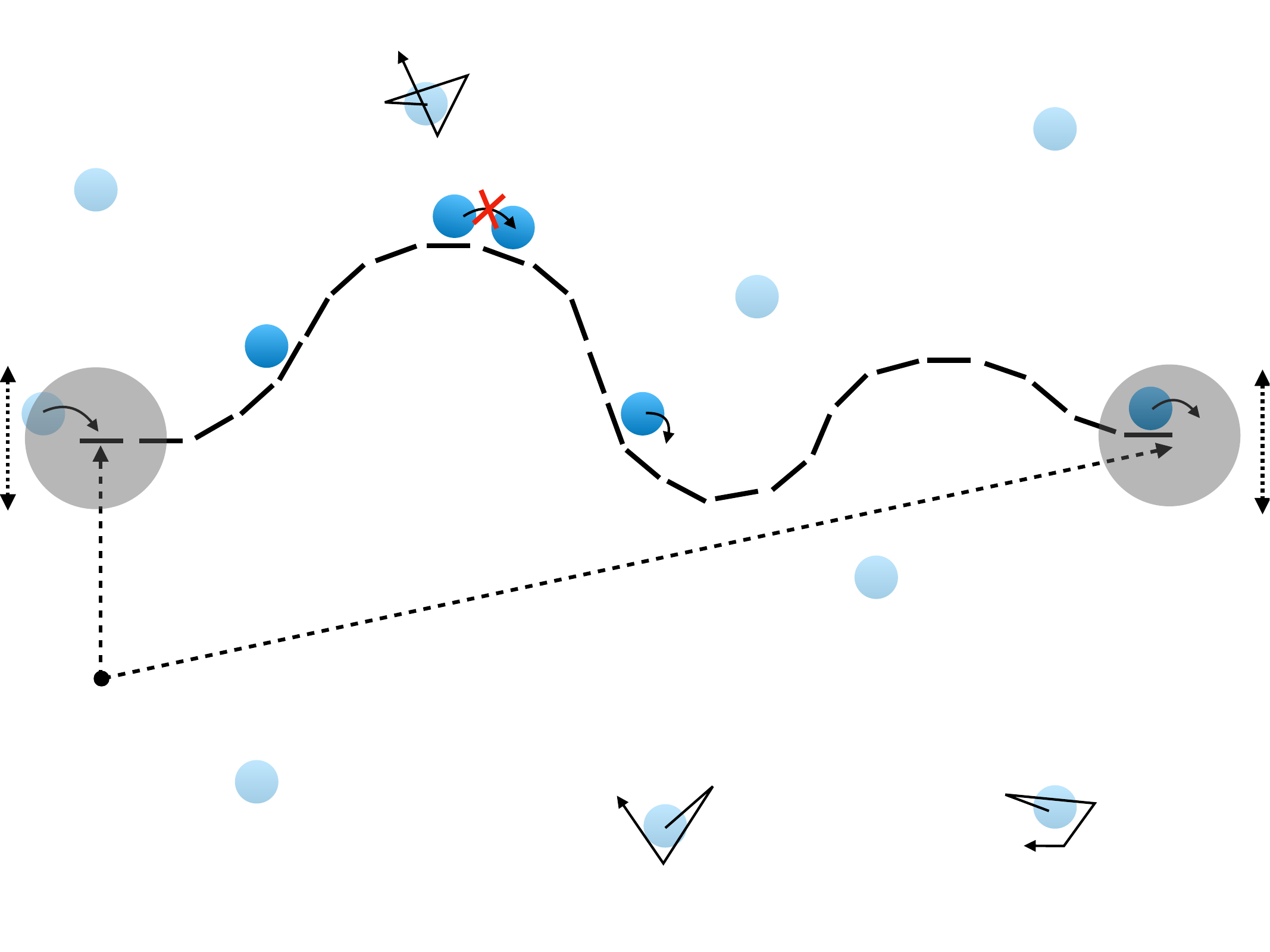}}
 \put(-17,117){\Large$\beta$}
 \put(4,100){\Large$2r$}
  \put(-275,100){\Large$2r$}
 \put(-232,70){\Large$\mathbf{r}_{\alpha}$}    
 \put(-140,65){\Large$\mathbf{r}_{\beta}$}
\put(-117,106){\Large$p$}
\put(-161,168){\Large$D$}
\put(-242,115){\Large$\tilde{\alpha}$}
\caption{{\it Graphical illustration of the Ribosome Transport with Diffusion model (RTD).}   The mRNA is  represented with a dashed line, ribosomes processing along the mRNA at a rate $p$ are represented by dark blue discs, and ribosomes diffusing freely at a diffusion coefficient $D$ are represented by light blue discs. Grey discs of radius $r$ centered at the end poinst of the mRNA  are the reaction volumes: if a diffusing ribosome is located in the reaction volume at the mRNA end-point centred around position $\bf r_\alpha$, then it attaches at a rate $\tilde\alpha$ to the mRNA.  On the other hand, if a ribosome is at the last site of the mRNA, then it detaches at a rate $\beta$ and is released inside the reaction volume centred around $\bf r_{\beta}$.}\label{figure1}
\end{figure}
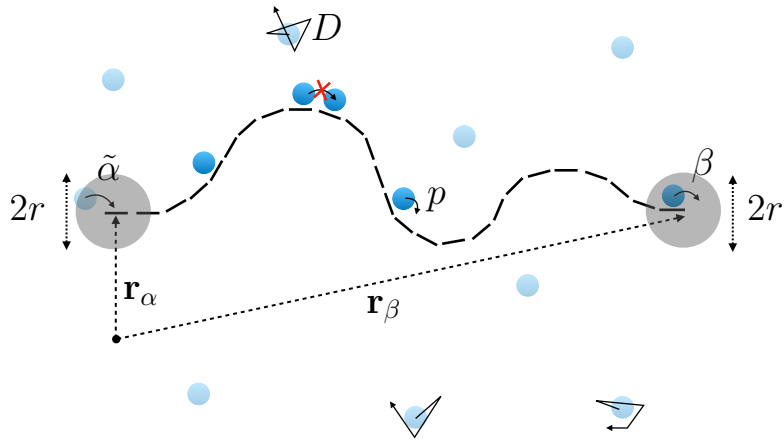

We introduce here the RTD, a minimalistic model that allows us to study how diffusion determines the rate of protein synthesis.   The RTD consists of ribosomes that diffuse in a  medium embedded in two or three dimensions and can bind to a one-dimensional substrate, say a mRNA filament. Bound ribosomes  then move unidirectionally along the filament by converting the intracellular chemical energy from the hydrolysis of guanine triphosphate (GTP) into mechanical motion, which is modelled by a Totally Asymmetric Simple Exclusion Process (TASEP).   In Fig.~\ref{figure1}, we present an illustration of the model and its parameters.  

%--------------------------------------------------------------------------------------------------
%--------------------------------------------------------------------------------------------------

We consider a  filament immersed in a medium containing ribosomes at a concentration  $c_{\infty}$. The filament  is a monopolymer consisting of $\ell$ monomers of length $a$. The first and last monomers of the filament are located at positions $\mathbf{r}_{\alpha}$ and  $\mathbf{r}_{\beta}$, respectively. For simplicity, we consider that $\mathbf{r}_{\alpha}$ and $\mathbf{r}_{\beta}$ are fixed in time.  
  
The dynamics of unbound molecular motors is modelled as a Brownian motion with diffusion coefficient $D$.

%In three dimensions, the diffusion coefficient takes the form 
%\begin{equation}
%D = \frac{k_{\rm B}T}{6\pi \eta r},  \label{eq:D3d}
%\end{equation}
%where $T$ is the absolute temperature, $k_{\rm B}$ is the Boltzmann's constant, $\eta$ is the viscosity of the medium, and $r$ is the radius of the ribosome. In two dimensions, the relation between diffusion and viscosity is more complicated than given by  Eq.~(\ref{eq:D3d}), see Ref.~\cite{Saf}. We assume that unbound ribosomes do not interact with each other.    

The dynamics of bound molecular motors is a unidirectional, hopping process with excluded volume interactions, which we model with a  TASEP on a one-dimensional lattice of length $L = \ell a$ \cite{Macdonald1968,Macdonald1969,Chou2011,Chow2013}. The TASEP model is a Markov jump process with the following rates: the hopping (or elongation) rate $p$ at which particles make a step of length $a$, the exit rate $\beta$ at which particles detach from the filament  end-point, and the entry rate  
\begin{equation}
\alpha(t)= \tilde{\alpha} \:N_{\rm r}(t), \label{eq:alpha}
\end{equation}
where $\tilde{\alpha}$ is the rate at which ribosomes contained in the reaction volume bind to the filament and $N_{\rm r}(t)$ is the number of ribosomes present in the reaction volume at time $t$. The reaction volume is considered to be a sphere (in three dimensions) or a disc (in two dimensions) of radius $r$ centered around the first monomer of the filament located at $\mathbf{r}_{\alpha}$. The reaction volume radius is of the same order of magnitude as the size of a ribosome. When ribosomes  detach from the filament they appear at a random location in a sphere (in three dimensions) or disc (in two dimensions) of radius $r$ centered around $\mathbf{r}_{\beta}$. Because of excluded volume interactions, each monomer can be bound to at most one ribosome. Therefore, ribosomes cannot hop forward if the subsequent monomer is already occupied by a ribosome and ribosomes cannot bind to the first monomer when it is already occupied, as illustrated in Fig.~\ref{figure1}.

%--------------------------------------------------------------------------------------------------    
\section{Mean-field theory for  coupling of  diffusion with active transport } \label{Sec:theory}  
%--------------------------------------------------------------------------------------------------

We present a mean field theory for the RTD model that couples diffusion with active transport. First, in Sec.~\ref{Sec:IIIA}, we  discuss how the protein synthesis rate is related to the stationary current of the TASEP model.   Second, in Sec.~\ref{Sec:IIIB}, we derive an analytical expression for the protein synthesis rate that is independent of the geometrical properties of the medium or reservoir in which the one-dimensional substrate is immersed, in the sense that the geometrical aspect of the problem is captured in the value of one nonuniversal constant. Lastly, in Sec.~\ref{Sec:IIIC}, we discuss the impact of the geometry of the surrounding reservoir on the protein synthesis rate.

% --------------------------------------------------------------------------------------
\subsection{Protein synthesis rate is given by the stationary current on the filament} \label{Sec:IIIA}
% --------------------------------------------------------------------------------------

The quantity of interest from a biological point of view is the protein synthesis rate $J$, which corresponds with the stationary current  of particles on the filament \cite{Macdonald1968, Macdonald1969}.    

The  stationary current of the RTD model in the limit of infinitely large $D$ is equal to the stationary current $\mathcal{J}$ of the TASEP model.   In the limit of 
large $\ell$, it holds that \cite{Derrida, Chou2011, Blythe}
\begin{eqnarray}
\mathcal{J} = \left\{\begin{array}{ccc}   \alpha \left(1-\frac{ \alpha }{p}\right),  &&\alpha<\beta \ {\rm and} \ \alpha <p/2,  \quad {\rm (LD)}, \\ \beta \left(1-\frac{\beta}{p}\right), && \beta<\alpha \ {\rm and} \  \beta <p/2,  \quad {\rm (HD)},  \\ \frac{p}{4}, &&   \alpha   \geq p/2 \ {\rm and} \ \beta \geq p/2,   \quad {\rm (MC)} . \end{array} \right. \label{eq:JTASEP}
\end{eqnarray}   
The three branches in Eq.~(\ref{eq:JTASEP}) correspond with three nonequilibrium phases: a Low-Density phase (LD) at small entry rates $\alpha<\beta$ and $ \alpha <p/2$, a High-Density phase (HD) at small exit rates $\beta<\alpha$ and $\beta<p/2$, and a Maximal Current phase (MC) when both  $ \alpha \geq p/2$ and $\beta\geq p/2$. In the LD phase, the ribosome attachment process is rate limiting and the current is a function of $\alpha $; in the HD phase, the ribosome detachment process is rate limiting and the current is a function of $\beta$; and in the MC phase, the filament hopping process is rate limiting and the current is independent of both $ \alpha$ and $\beta$. Experimental data in yeast cells \cite{MacKay}  and in neurons of mammals \cite{Biever} show that  the rate limiting process  for translation is the initiation of ribosomes.    

In the RTD model at finite values of $D$, the entry rate $\alpha(t)$ on the filament is not a constant but a fluctuating quantity, see Eq.~(\ref{eq:alpha}). In the stationary state,  the average  current $J$ is well approximated  by the expression (\ref{eq:JTASEP}) with the entry rate $\alpha$ replaced by its average value 
 \begin{eqnarray}
 \langle \alpha(t) \rangle= \tilde{\alpha} \: \langle N_{\rm r}(t) \rangle \,,
 \end{eqnarray}
 where $\langle \cdot \rangle$ denotes the average over many realizations of the stationary process.
Since in the stationary state the average number $\langle N_{\rm r}(t) \rangle$ of ribosomes in the reaction volume is independent of time, we set
   \begin{eqnarray}
    \langle \alpha(t) \rangle=  \langle \alpha \rangle.
  \end{eqnarray}
Replacing in Eq.~(\ref{eq:JTASEP})  $\alpha$ by $\langle \alpha \rangle$, we obtain for the stationary current of the RTD model the expression
 \begin{eqnarray}
J =  \left\{\begin{array}{ccc}  \langle \alpha \rangle\left(1-\frac{\langle \alpha \rangle}{p}\right),  && \langle \alpha \rangle < \beta \ {\rm and}\ \langle \alpha\rangle < p/2,   \quad {\rm (LD)}, \\ \beta \left(1-\frac{\beta}{p}\right), && \beta<    \langle \alpha \rangle  \ {\rm and}\ \beta< p/2, \quad {\rm (HD)},  \\ \frac{p}{4}, &&  \langle \alpha \rangle \geq p/2 \ {\rm and} \  \beta \geq \frac{p}{2},   \quad {\rm (MC)} . \end{array} \right. \label{eq:J}
\end{eqnarray} 
Note that replacing $\alpha$ by $\langle \alpha\rangle$ is a mean-field approximation because it neglects correlations between particles in the reaction volume and particles on the filament.
From Eq.~(\ref{eq:J}) we observe that if the filament is in the HD or MC phase, then the protein synthesis rate  is independent of the diffusion process in the reservoir. However, in the LD phase when the initiation step is rate limiting, which is biologically relevant case, the current $J$ depends on  the concentration of unbound ribosomes through $\langle \alpha\rangle$, and hence in this regime we are required to include diffusion into  our theoretical analysis.   Often it will be insightful to consider the limiting case where particle excluded volume on the filament is irrelevant for which  the  simpler formula
\begin{eqnarray}
J=\langle \alpha\rangle \label{eq:JnoExcl}
\end{eqnarray} 
holds. Note that this condition is fulfilled for low density of ribosomes on the filament.

%-----------------------------------------------------------------------
\subsection{Protein synthesis rate: universal expression}\label{Sec:IIIB}
%-----------------------------------------------------------------------

From the point of view of the reservoir of diffusing ribosomes the filament serves  both as a sink  and a source of ribosomes. 

If the initiation and termination sites overlap, as will be approximately the case for  circular mRNA, then the concentration of ribosomes in the reservoir will be homogeneous since  source and sink  exactly compensate for each other, and therefore  in this case
\begin{eqnarray}
\langle \alpha \rangle = \alpha_{\infty} = \tilde{\alpha}c_{\infty}\mathcal{V}\,,   \label{Eq:AlphaInf}
\end{eqnarray}
where $\mathcal{V}$ is the reaction volume of radius $r$, which for two dimensions and three dimensions is given by  $\mathcal{V} = \pi r^2$  and $\mathcal{V} = 4\pi r^3/3$, respectively.

On the other hand, if the termination site is distant from the initiation site, then $\langle \alpha\rangle$ will have a reduced value, with respect to Eq.~(\ref{Eq:AlphaInf}) due  to the depletion of ribosomes in the reaction volume at the initiation site.   Indeed,  the current on the filament carries away ribosomes from the reaction volume, which in the stationary state will be compensated by the diffusive current in the reservoir. As we will show in the next section, the depletion effects due to finite  diffusion are captured by the formula 
\begin{eqnarray}
\langle \alpha \rangle
 &=& \alpha_{\infty} \left(1 - \frac{J\mu_{d}}{D_{\rm eff}\alpha_{\infty}}\right) ,  \label{eq:alpha3D}
\end{eqnarray}   
where $\mu_d$ is a constant that depends on the geometry of the problem and where
\begin{eqnarray}
D_{\rm eff} = \frac{D}{\tilde{\alpha}r^2} \label{eq:DEff}
\end{eqnarray} 
is an effective diffusion coefficient. The dimensionless quantity $D_{\rm eff}$ quantifies the competition between injection of ribosomes on the filament and the diffusion of ribosomes into the reaction volume. Equation (\ref{eq:alpha3D}) follows from solving the diffusion equation for ribosomes in the reservoir, as we shall describe in detail in the next section. Eq.~(\ref{eq:alpha3D}) states that the  rate $\langle \alpha \rangle$ is  the sum of the entry rate $\alpha_{\infty}$ for a homogeneous reservoir minus a correction term that captures the effect of  finite diffusion on the entry rate.  The correction term is negative since the filament depletes particles in the reaction volume at the initiation site. Moreover,  Eq.~(\ref{eq:alpha3D}) states that  the correction term is proportional to the current $J$ on the filament, inversely proportional to the effective diffusion constant $D_{\rm eff}$, and it is also proportional to  the dimensionless, nonuniversal constant $\mu_d$ that depends, as we shall see in the next section, on the geometrical properties of the system, namely,   the end-to-end distance $|\mathbf{r}_{\beta}-\mathbf{r}_{\alpha}|$,   the location of the filament in the reservoir, the dimensionality of the system, and the boundary conditions of the reservoir of diffusing ribosomes.  Here, we would like to focus on the physical consequences of the Eq.(9).
%We will compute $\mu_d$ on several examples in the next section, while in the remaining part of this section we analyse the physical consequences of the Eq.~(\ref{eq:alpha3D}) for the protein synthesis rate $J$.

% ---------------------------------------------------------------------------------------------------
\begin{figure}
{\includegraphics[width=0.4\textwidth]{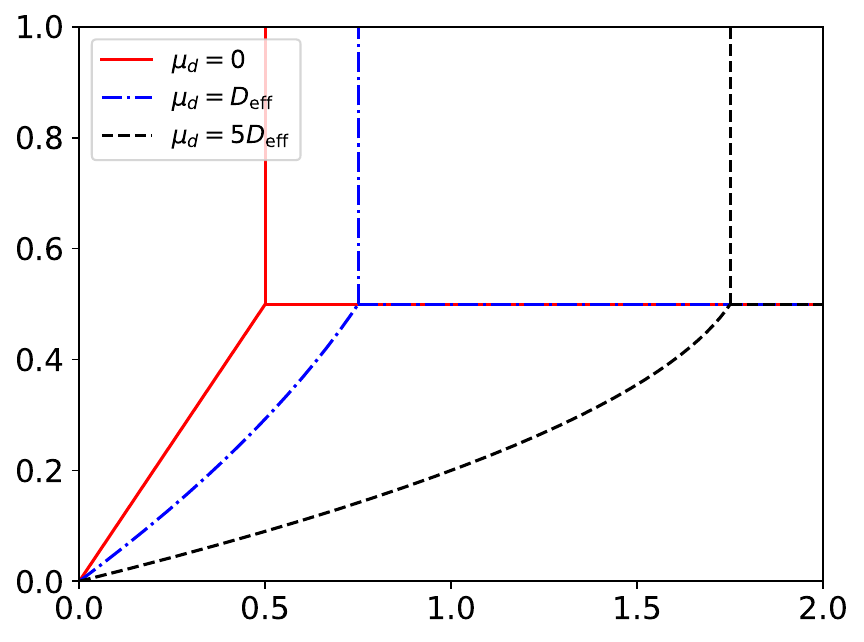}}
\hspace{4mm}
{\includegraphics[width=0.4\textwidth]{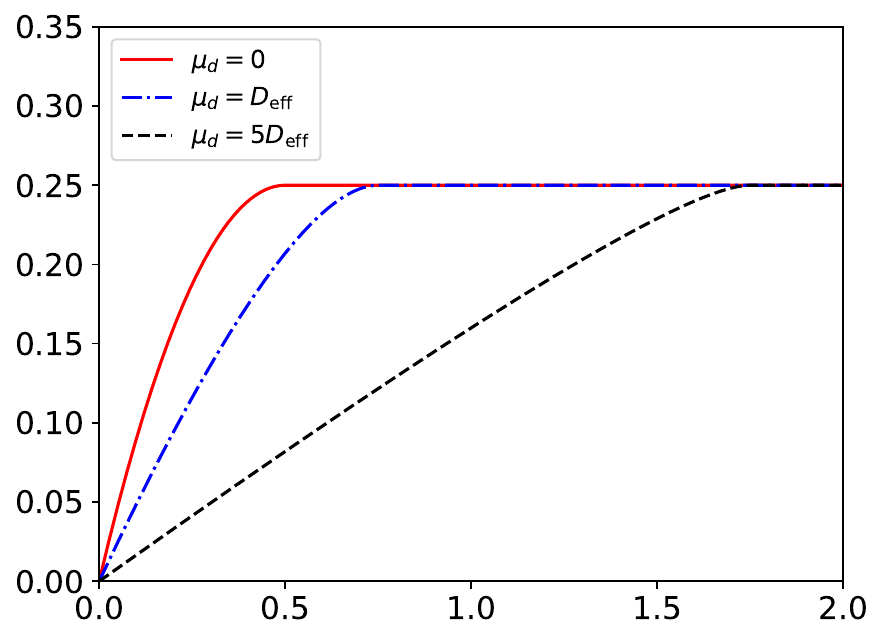}}
 \put(-390,90){\Large \color{black}LD}
  \put(-265,40){\Large \color{black}HD}
    \put(-253,95){\Large \color{black}MC}
     \put(-445,80){\large $\beta/p$} 
     \put(-335,-10){\large $\alpha_{\infty}/p$}
          \put(-110,-10){\large $\alpha_{\infty}/p$}
               \put(-223,80){\large $J/p$} 
 \put(-250,125){\Large (a)}
  \put(-30,125){\Large (b)}
\caption{Panel (a): phase diagram for the RTD model for three values of the parameter $\mu_d/D_{\rm eff}$. Panel (b): protein synthesis rate $J/p$ in the RTD model as a function of the ratio $\alpha_{\infty}/p$ for a large exit rate $\beta>p/2$.}\label{phase_diagram}
\end{figure}
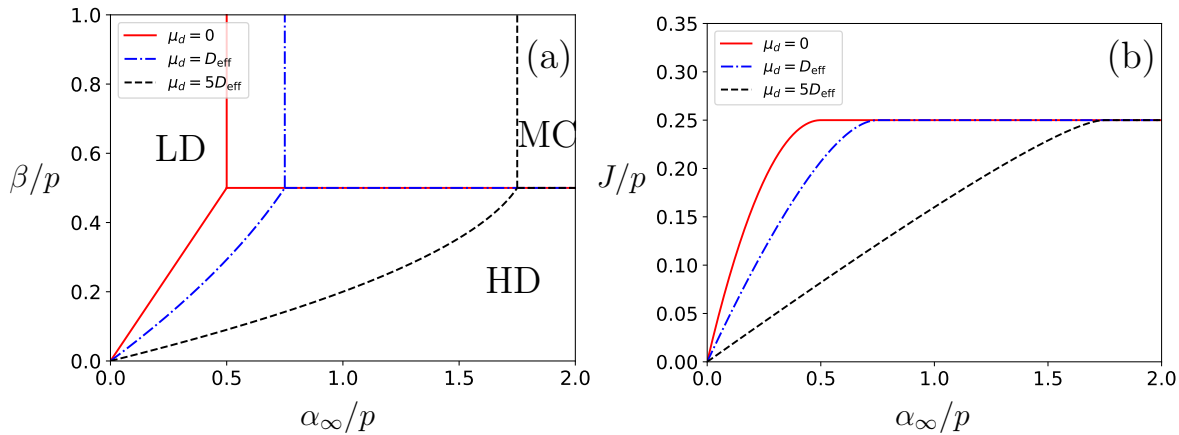
%-----------------------------------------------------------------------------------------------------

To obtain the protein synthesis rate $J$, we combine  Eqs.~(\ref{eq:J}) and~(\ref{eq:alpha3D}). In the  LD phase, we obtain a second-order algebraic equation whose solution  $\langle \alpha \rangle\in [0,p/2]$ is given by
 \begin{eqnarray} 
 \langle\alpha\rangle=p\frac{D_{\rm eff}+\mu_{d}}{2\mu_{d}}\left(1-\sqrt{1-4\zeta}\right) \,,\label{eq:alphaFinal}
\end{eqnarray}
where the adimensional parameter 
\begin{equation}
\zeta=\frac{\alpha_\infty D_{\rm eff}\mu_{d}}{p(D_{\rm eff}+\mu_{d})^2} \label{eq:zeta}
\end{equation}
quantifies the effect of exclusion on $\langle \alpha \rangle$. The argument of the square root in (\ref{eq:alphaFinal}) is always positive when the filament is in the LD phase because in the LD phase $\langle \alpha \rangle=p\frac{D_{\rm eff}+\mu_{d}}{2\mu_{d}}<p/2$, which implies $\zeta<1/4$.  Note that if the  diffusion coefficient $D_{\rm eff}$ is small enough, then $\zeta\ll1$ and exclusion has a minor effect. Plugging $\langle\alpha\rangle$ inside the expression for the current, given by Eq.~(\ref{eq:J}), we obtain the following expression for the protein synthesis rate, 
 \begin{eqnarray}
J =  \left\{\begin{array}{ccc}  \langle\alpha\rangle\left(1- \langle\alpha\rangle/p\right),  &&
\alpha_{\infty}<\beta \left[1 + \frac{\mu_{d}}{D_{\rm eff}}(1-\beta/p)\right] \ {\rm and}\ \alpha_{\infty}< p/2\left(1 + \frac{\mu_d}{2D_{\rm eff}}\right), 
  \quad {\rm (LD)}, \\ \beta \left(1-\frac{\beta}{p}\right), && \alpha_{\infty}>\beta \left[1 + \frac{\mu_{d}}{D_{\rm eff}}(1-\beta/p)\right] \ {\rm and}\ \beta< p/2 , 
  \quad {\rm (HD)},  \\ \frac{p}{4}, &&  \alpha_{\infty}\geq p/2\left(1 + \frac{\mu_d}{2D_{\rm eff}}\right)\ {\rm and}\ \beta\geq p/2 , 
  \quad {\rm (MC)} , \end{array} \right. \label{eq:J3D}
\end{eqnarray}
where $ \langle\alpha\rangle$  is given by  (\ref{eq:alphaFinal}). For small values of  $\zeta$, the filament will be in the LD phase, and we obtain the simpler expression 
\begin{eqnarray}
J = \frac{\alpha_{\infty}  D_{\rm eff}}{D_{\rm eff}+\mu_d},   \label{noExclu}
\end{eqnarray}  
which also follows from Eq.~(\ref{eq:JnoExcl}).    Equation (\ref{eq:J3D}) implies that the current $J$ admits a universal expression that only depends on four parameters: the entry rate $\alpha_{\infty}$ for a homogeneous reservoir,  the elongation rate $p$, the exit rate $\beta$, and the parameter $\mu_d/D_{\rm eff}$ that quantifies the effect of finite diffusion on the current $J$. From Eqs.(\ref{eq:J3D}) and (\ref{noExclu}) it also follows that the effect of finite mobility of ribosomes on the protein synthesis rate $J$ is significant when  $\mu_d\gg D_{\rm eff}$. On the other hand, when $\mu_d\ll D_{\rm eff}$, then the finite mobility of ribosomes will be irrelevant for $J$.   

 In Fig.~\ref{phase_diagram}(a), we present the  phase diagram for the RTD model for three values of $\mu_d/D_{\rm eff}$, namely, the case with an infinite diffusion rate, $\mu_d/D_{\rm eff}=0$, and two cases with finite diffusion rates, $\mu_d=D_{\rm eff}$ and $\mu_d=5D_{\rm eff}$.   For $\mu_d/D_{\rm eff}=0$, we recover the  phase diagram of TASEP \cite{Derrida, Chou2011, Blythe}, while for finite values of $\mu_d$ we observe an increase of the      LD phase and a corresponding decrease of the MC and HD phases.  This is because finite diffusion depletes particles in the reaction volume surrounding the initiation site of the filament, and hence reduces the current on the filament for a given $\alpha_\infty$. This is shown in Fig.~\ref{phase_diagram}(b), where we plot the current as a function of $\alpha_\infty/p$ for fixed a  value of $\mu_d/D_{\rm eff}$ and $\beta/p\geq 1/2$. If $\mu_d\ll D_{\rm eff}$, then the reservoir is homogeneous and we obtain the standard TASEP result    \cite{Derrida, Chou2011, Blythe}
\begin{eqnarray}
J =  \left\{\begin{array}{ccc} \alpha_{\infty}(1-\alpha_{\infty}/p),  && \alpha_{\infty}<p/2 ,\\  p/4,  && \alpha_{\infty}>p/2.\end{array}\right. \label{eq:TASEPlimit}
\end{eqnarray}
In the opposing limiting case when $\mu_d \gg  D_{\rm eff} $ the reservoir is strongly inhomogeneous and we obtain that
\begin{eqnarray}
J =  \left\{\begin{array}{ccc}\frac{D_{\rm eff}\alpha_{\infty}}{\mu_{d}} ,  && \alpha_{\infty}<p \mu_d/4 ,\\  p/4,  && \alpha_{\infty}>p\mu_d/4.\end{array}\right.
\end{eqnarray}   
  In this limit the  environment is viscous and therefore the effects of excluded volume become negligible.
  
Note that the results of Fig.2 do not consider the effects of finite resources. Therefore, it is implicitely assumed that the number of ribosomes is very large compared to the average number of ribosomes on the mRNA. In the case of finite resources, the phase diagram displays an extended shock phase, as shown in Refs.~\cite{Dauloudet,Cook09}.   

So far, much of the interesting physics has been hidden in the dimensionless constant  $\mu_d$ that depends on the geometry of the problem. In the next subsection we will explicitly solve the diffusion equation coupled to directed transport on the filament to obtain explicit expressions for $\mu_d$.

\begin{figure}
{\includegraphics[width=0.5\textwidth]{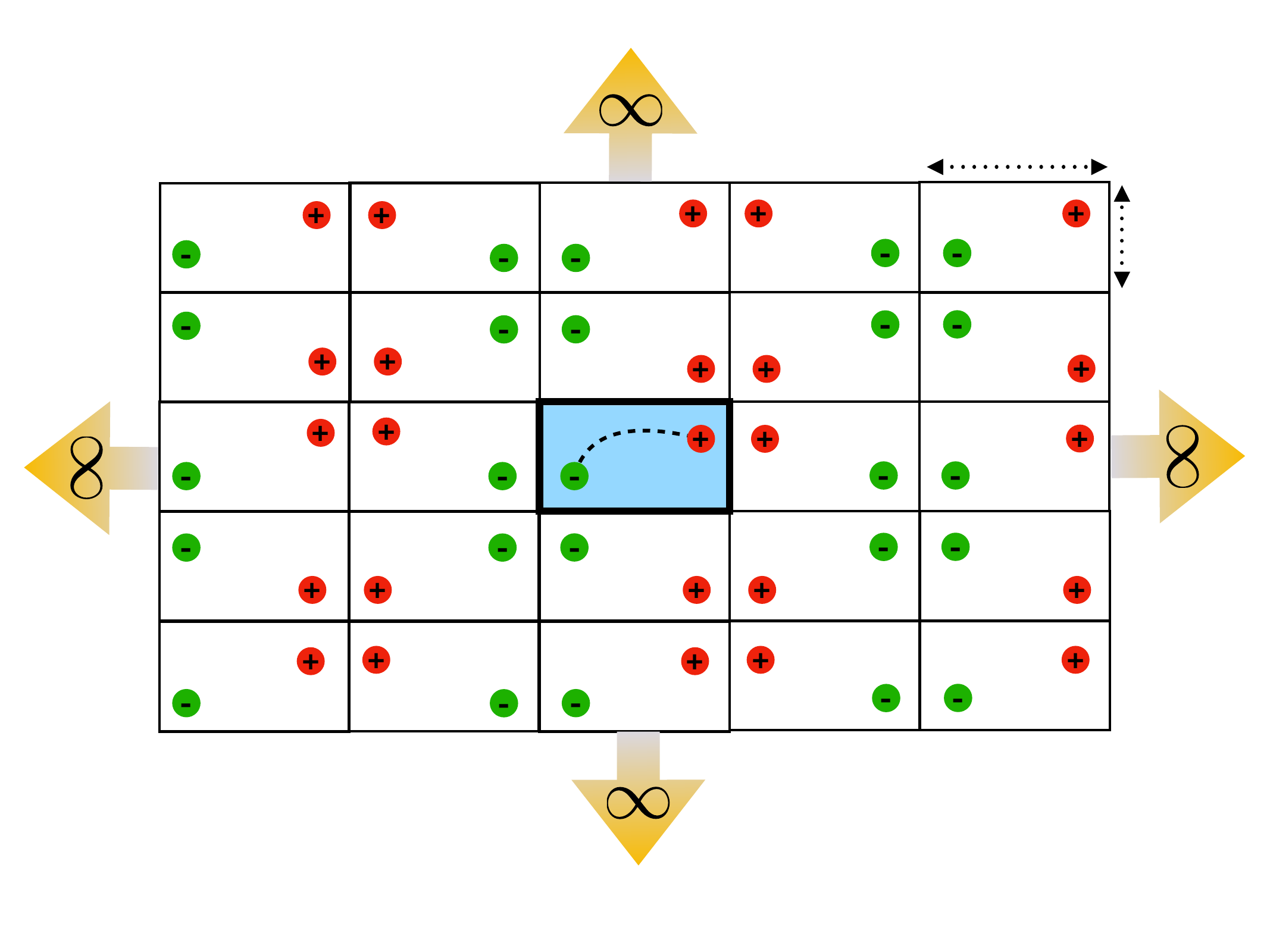}}
 \put(-53,163){$L_x$}
  \put(-25,143){$L_y$}
\caption{{\it Illustration of the method of images}:   Diffusion of ribosomes in a confined rectangular  box is  equivalent to diffusion  of ribosomes in a two-dimensional Euclidean space that contains an infinite number of images of the original source (denoted by red) and  sink (denoted by green) located in the rectangular  box (located in the center and coloured in blue).}\label{fig:image}
\end{figure}

% ------------------------------------------------------------------------------------
\subsection{Influence of geometry on the protein synthesis rate}\label{Sec:IIIC}
% ------------------------------------------------------------------------------------

In order to obtain an expression for $\mu_d$, and thus  complete the theoretical treatment for ribosomes with finite mobility, we solve the diffusion equation in the reservoir coupled with active transport on the filament.   We consider the case where $|\mathbf{r}_{\beta}-\mathbf{r}_{\alpha}|>2r$ so that the reaction volumes at the source and the sink do not overlap.

The concentration $c(\mathbf{r},t)$ of unbound ribosomes  at the spatial coordinate $\mathbf{r}\in\mathbb{R}^d$ and at the time $t$ is described by the diffusion equation:
\begin{equation}
 \frac{\partial c(\mathbf{r},t)}{\partial t}= -\vec\nabla\cdot\vec J_d(\mathbf{r},t)-\Pi(\mathbf{r},t)\,,
\end{equation}
where
\begin{equation}
 \vec J_d(\mathbf{r},t)=-D\vec\nabla c(\mathbf{r},t)
\end{equation}
is the diffusive current, and 
\begin{equation}
 \Pi(\mathbf{r}) =  \left\{\begin{array} {ccc}  \frac{J}{\mathcal{V}}&& |\mathbf{r}-\mathbf{r}_{ \alpha}|\leq r, \\ -\frac{J}{\mathcal{V}}&& |\mathbf{r}-\mathbf{r}_{ \beta}|\leq r,  \\ 
 0 && |\mathbf{r}-\mathbf{r}_{\alpha}|>r \quad {\rm and\quad} |\mathbf{r}-\mathbf{r}_{\beta}|>r , \label{eq:PI}
 \end{array} \right.
\end{equation}  
is proportional to the rate $J$ at which particles exit and enter the filament. We have used that $|\mathbf{r}_{\beta}-\mathbf{r}_{\alpha}|>2r$.
The stationary concentration $c(\mathbf{r})$ of unbound ribosomes solves
\begin{equation}
 D \:\Delta  c(\mathbf{r}) = \Pi(\mathbf{r}),  \quad  \label{eq:Diffusion}
\end{equation}
where $\Delta$  is the Laplacian with respect the radius $\mathbf{r}$.

The diffusion equation admits the solution
\begin{eqnarray}
c(\mathbf{r}) = c_{\infty} + \int_{\mathbb{R}^d}{\rm d}^d\mathbf{r'}\,\Pi(\mathbf{r}')\: \mathcal{G}_{d}(\mathbf{r},\mathbf{r}'), \label{eq:c}
\end{eqnarray} 
where $\mathcal{G}(\mathbf{r},\mathbf{r}')$ is the Green function that solves 
\begin{eqnarray}
D\Delta \mathcal{G}_d(\mathbf{r},\mathbf{r}') = \delta(\mathbf{r}-\mathbf{r}').\label{eq:DeltaG}
\end{eqnarray}    
The entry rate $\langle \alpha \rangle$ is related to the stationary concentration in the reaction volume through 
  \begin{eqnarray}
\langle \alpha \rangle =  \tilde{\alpha} \int_{|\mathbf{r}-\mathbf{r}_{\alpha}|\leq r} c(\mathbf{r}) {\rm d}\mathbf{r} .  \label{eq:alphaV}
\end{eqnarray} 
Note that the latter equation is consistent with Eq.(3) because at the stationary state $\langle N_r(t)\rangle=\int_{|\mathbf{r}-\mathbf{r}_{\alpha}|\leq r} c(\mathbf{r}) {\rm d}\mathbf{r}$.

The explicit form of the Green's function and thus $\langle \alpha \rangle$ depend on the geometry of the reservoir. We provide below a couple of examples.

% ------------------------------------------------------------------------------------
\subsubsection{RTD in two-dimensional infinite box $(\mathbb{R}^2)$}
% ------------------------------------------------------------------------------------

In two dimensions, the Green function takes the form~\cite{Bressloff,Jackson} 
\begin{eqnarray}
\mathcal{G}_2(\mathbf{r},\mathbf{r}')  = -\frac{1}{2\pi D}\ln |\mathbf{r}-\mathbf{r}'| \label{eq:2dG} .
\end{eqnarray}
Substituting the  Green function in Eq.~(\ref{eq:c}), we obtain an explicit expression for $c(\mathbf{r})$, see Appendix~\ref{appendixA}.   Subsequently, substituting the explicit solution for $c(\mathbf{r})$ in Eq.~(\ref{eq:alphaV}) we obtain the formula Eq.~(\ref{eq:alpha3D}) with 
\begin{eqnarray}
\mu_{2} = \frac{ \log d_{\alpha\beta}+1}{2}, \label{eq:mu2x}
\end{eqnarray}
where 
\begin{eqnarray}
 d_{\alpha\beta} = \frac{|\mathbf{r}_{\beta}-\mathbf{r}_{\alpha}|}{r}  \label{eq:dalphabeta}
\end{eqnarray}
is the effective distance between the initiation site and the termination site on the filament.   Substitution of $\mu_d$ into Eqs.~(\ref{eq:alphaFinal}-\ref{eq:J3D}) provides us with an explicit expression for the current $J$ as a function of  $d_{\alpha\beta}$.

In Fig.~\ref{phase_diagram2}, we plot the current $J$ as a function of the separation $d_{\alpha\beta}$ between the two end-points of the mRNA for   two values of the effective diffusion constant $D_{\rm eff}$.   Although   the part for  $d_{\alpha\beta}<2$ is not covered by our calculations, we know that $J = \alpha_{\infty}(1-\alpha_{\infty}/p)$ for $d_{\alpha\beta}=0$, which in Fig.~\ref{phase_diagram2} corresponds to $J=0.24p$.  
  We observe that the current decreases monotonically as function of  $d_{\alpha\beta}$ and approaches zero for $d_{\alpha\beta}$ large enough.    The decay towards zero is logarithmically slow after a fast initial decay in the regime $d_{\alpha\beta}<2$ where initiation and termination sites overlap.    

% ------------------------------------------------------------------------------------
\subsubsection{RTD in three-dimensional infinite box $(\mathbb{R}^3)$} %-----------------------------------------------------
% ------------------------------------------------------------------------------------

In three dimensions the Green function is given by 
\begin{eqnarray}
\mathcal{G}_3(\mathbf{r},\mathbf{r}')  =  \frac{1}{4\pi D} \frac{1}{|\mathbf{r}-\mathbf{r}'|}. \label{eq:3dG}
\end{eqnarray}
Using this expression for the  Green function in Eq.~(\ref{eq:c}), we obtain an explicit expression for $c(\mathbf{r})$, see Appendix~\ref{appendixA}, which we substitute in Eq.~(\ref{eq:alphaV}) to obtain  formula Eq.~(\ref{eq:alpha3D}) with now
\begin{eqnarray}
\mu_{3} =  \frac{2}{5} - \frac{1}{3d_{\alpha\beta}}. \label{eq:mu3x}
\end{eqnarray}
Comparing Eqs.~(\ref{eq:mu2x}) and (\ref{eq:mu3x}), we see that there is a  difference between two and three dimensions: in three dimensions $\mu_3$ converges to a finite value for $d_{\alpha\beta} \rightarrow \infty$ whereas in two dimensions $\mu_2$ diverges for $d_{\alpha\beta} \rightarrow \infty$.    This implies that in two dimensions $J$ converges to zero for large  distances $d_{\alpha\beta}$ between the end-points of the filament, while it converges to a finite nonzero value in three dimensions. 

The distinction between the dependency of the current $J$ in two and three dimensions is illustrated in Fig.~\ref{phase_diagram2}.   In three dimensions, the current saturates fast to its asymptotic value after an initial quick decay for values  $d_{\alpha\beta}<2$.      The asymptotic value  of $J$ depends on the diffusion constant $D_{\rm eff}$ and decreases to zero for $D_{\rm eff}\rightarrow 0$.    Hence, in three dimensions, the mRNA will carry a finite current, even when $d_{\alpha\beta}\rightarrow \infty$, and this asymptotic current will depend on the diffusion constant.    

In Fig.~\ref{currentDiffusion}, we plot the asymptotic current $J$ as a function of the effective diffusion constant $D_{\rm eff}$. We observe from Fig.~\ref{currentDiffusion} that at finite $D_{\rm eff}$ the protein synthesis rate in $d=2$ dimensions is smaller than the synthesis rate in $d=3$ dimensions. This is  because diffusive currents are smaller in lower dimensions and hence ribosomes are more depleted at the filament entrance.   For small values of $D_{\rm eff}$, the current is proportional to $D_{\rm eff}$, namely, 
\begin{equation}
J= \frac{\alpha_{\infty}}{\mu_d}D_{\rm eff}+ O(D^2_{\rm eff}),
\end{equation}
where the  proportionality constant is the  ratio between the entry rate $\alpha_{\infty}$  for  circularized mRNA and the constant $\mu_d$ that depends on the geometry of the problem.  

% ------------------------------------------------------------------------------------
\subsubsection{Two-dimensional rectangular box}
% ------------------------------------------------------------------------------------ 
Since the volume of a cell is  finite, it is relevant to understand  how the confinement of the mRNA in the cell affects the protein synthesis rate.

We first consider the case of a filament immersed into a medium that has the shape of a two-dimensional rectangular box.    We assume that the box is centered at the origin $\mathbf{r} = 0$ and that the sides of the box have lengths $L_x$ and $L_y$.  

We derive an explicit expression for the  Green function  in a two-dimensional rectangular box with the method of images~\cite{Feynman}.  The Green function of a point source  in a two-dimensional rectangular box  is identical to a series of  Green functions in $\mathbb{R}^2$ associated with   images of the point source, namely, it holds that 
\begin{eqnarray}
\mathcal{G}_{L_x,L_y}(\mathbf{r},\mathbf{r}')  = \mathcal{G}_2(\mathbf{r},\mathbf{r}') + \sum_{j\in \mathcal{N}} \mathcal{G}_2(\mathbf{r},\mathbf{r}^{(j)}), \label{eq:2dG-box}
\end{eqnarray} 
where $\mathbf{r}^{(j)}$ are the coordinates for the images  of the point source located  at $\mathbf{r}'$, see Fig.~\ref{fig:image} for an example, and $\mathcal{G}_2$ is the Green function in Eq.~(\ref{eq:2dG}). 

Substituting the Green function given by Eq.~(\ref{eq:2dG-box}) in Eq.~(\ref{eq:alphaV}), we obtain the expression Eq.~(\ref{eq:alpha3D}), with  now
 \begin{eqnarray}
 \mu_2(L_x,L_y)= \frac{ 1 + \log d_{\alpha\beta} + \mathcal{I}_{L_x, L_y} }{2},
 \end{eqnarray}  
 and where $\mathcal{I}_{L_x, L_y}$ is the series 
 \begin{eqnarray}
 \mathcal{I}_{L_x, L_y} = \sum_{j\in \mathcal{N}_{\beta}} \log  |\mathbf{r}_{\alpha}-  \mathbf{r}^{(j)}_{\beta}| - \sum_{j\in \mathcal{N}_{\alpha}} \log  |\mathbf{r}_{\alpha}-  \mathbf{r}^{(j)}_{\alpha}|  .   \label{eq:sums}
 \end{eqnarray} 
The sums in Eq.~(\ref{eq:sums}) run over the images of the initiation and termination sites of the filament, which defines the set $\mathcal N_\alpha$ and $\mathcal N_\beta$. The specific locations of $\mathbf{r}^{(j)}_{\alpha}$ and $\mathbf{r}^{(j)}_{\beta}$ are detailed in Fig.~\ref{fig:image}.    As shown in  Ref.~\cite{Dauloudet},  the series Eq.~(\ref{eq:sums}) converges rapidly since the influence of the  copies $\mathbf{r}^{(j)}_{\alpha}$ and $\mathbf{r}^{(j)}_{\beta}$ on the concentration of ribosomes in the original system decreases fast enough with the distance.

%Note that since the mean-field theory in this section applies to the case of a very large number of ribosomes in the reservoir, the concentration of ribosomes $c_\infty$ is also very large. Thus, in order to keep a physiological value of $\alpha_\infty$, as defined in Eq.(8), we need to take $\tilde\alpha$ small enough.
Note that the method of images  also works for a  triangular or hexagonal shaped cell as two-dimensional Euclidean space can be tiled with triangles and hexagons, see~\cite{Dauloudet} and references therein.     

% ------------------------------------------------------------------------------------
\subsubsection{Three-dimensional cuboid} %--------------------------------------------
% ------------------------------------------------------------------------------------
Since a cell is three dimensional, we consider now the case of a three-dimensional cuboid  with linear dimensions $L_x$,  $L_y$ and $L_z$.

An analytical expression for the protein synthesis rate can also be derived in the case of a  cuboid. We then obtain formula Eq.~(\ref{eq:alpha3D}) with 
 \begin{eqnarray}
 \mu_3(L_x, L_y,L_z) = \frac{2}{5} - \frac{ 1}{3} \left( \frac{1}{d_{\alpha\beta}} + \mathcal{I}_{L_x,L_y,L_z}\right)
 \end{eqnarray} 
 where $\mathcal{I}_{L_x,L_y,L_z}$ is the series 
 \begin{eqnarray}
 \mathcal{I}_{L_x,L_y,L_z}= \sum_{j\in \mathcal{N}_{\beta}}  \frac{r}{ |\mathbf{r}_{\alpha}-  \mathbf{r}^{(j)}_{\beta}|} - \sum_{j\in \mathcal{N}_{\alpha}} \frac{r}{ |\mathbf{r}_{\alpha}-  \mathbf{r}^{(j)}_{\alpha}| } .
 \end{eqnarray} 
The sums run over the images of the initiation and termination sites of the filament in $\mathbb{R}^3$. 

In Fig.~\ref{currentDist} we plot the protein synthesis rate $J$ as a function of  the height of the cuboid $L_z$ while keeping $\alpha_{\infty}$  fixed.  We observe that confinement reduces the current on the filament: the filament current in a confined volume with finite $L_z$ is smaller than one would expect for  $L_z=\infty$.  In addition, we observe that the effect of confinement is negligable when $L_z>20r$ with $r$ the radius of the reaction volume at the first site of the filament.

It will be interesting to extend the analysis to the case of a spheroid or cylinder.  Unfortunately, the method of images does not work for such volumes and therefore we need to resort to other mathematical techniques in order to solve diffusion in such volumes [Ref.???]. Since  the main effect of confinement is the reduction of the volume, one can use the results for a three-dimensional cuboid to estimate the overall influence of confinement on protein synthesis rates, even for cells with a spheroid or cylindrical shape.

% -------------------------------------------------------------------------------------
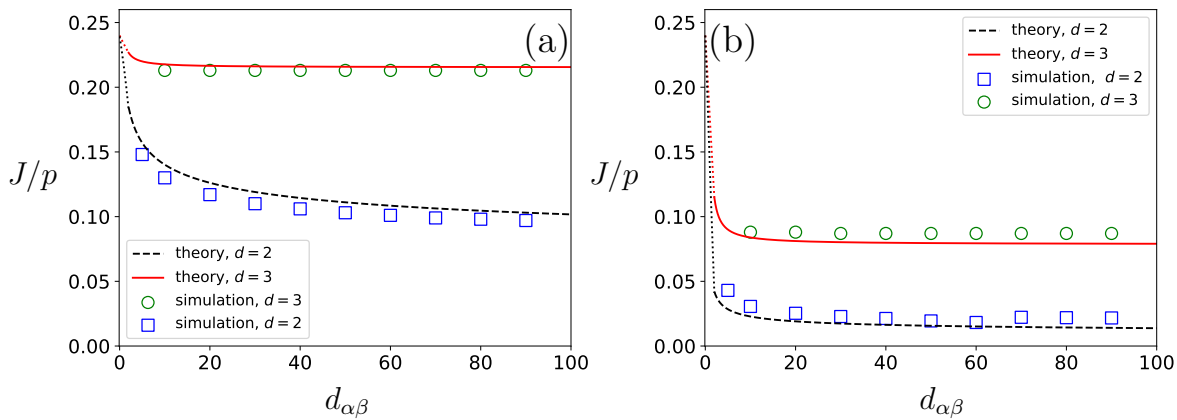
\begin{figure}
{\includegraphics[width=0.4\textwidth]{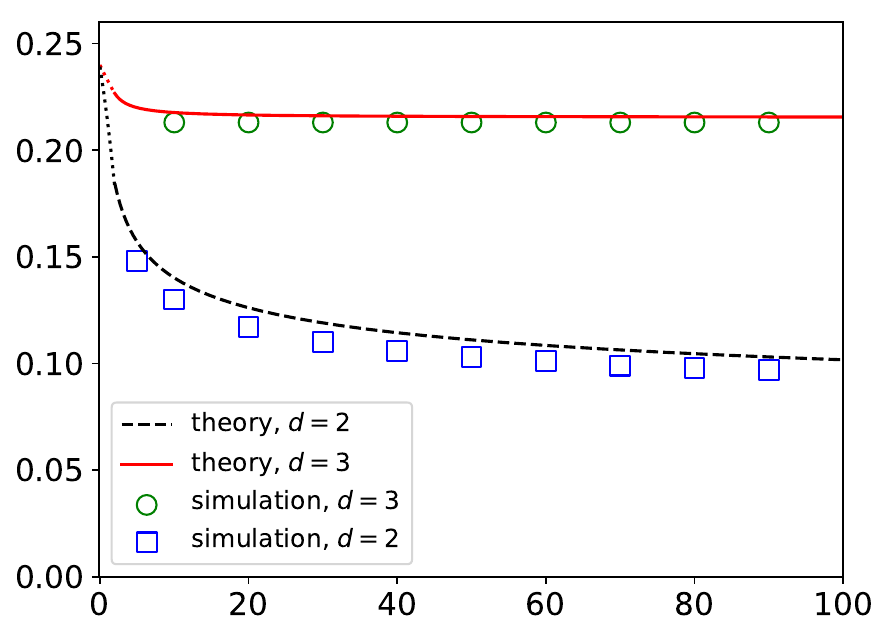}}
\hspace{4mm}
{\includegraphics[width=0.4\textwidth]{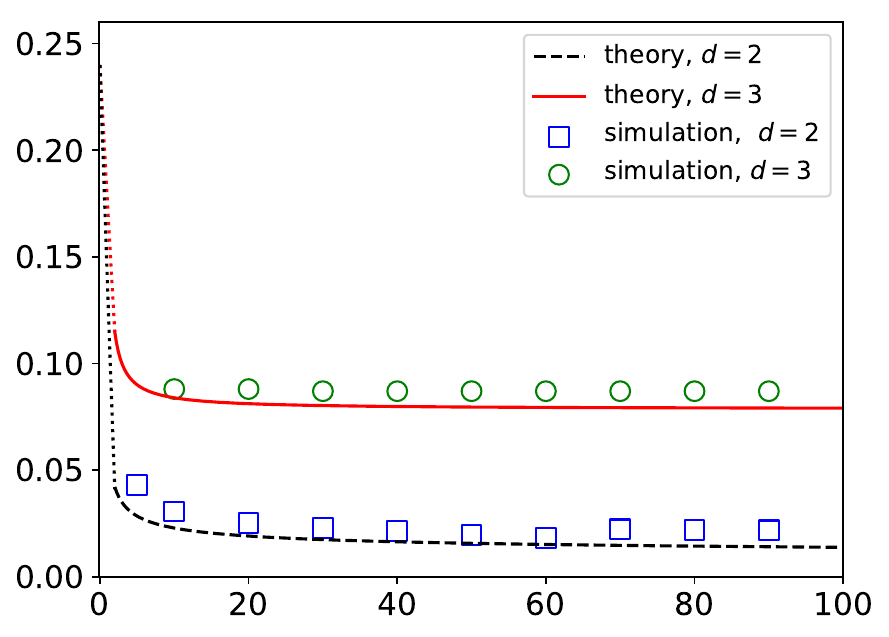}}
 \put(-445,75){\large$J/p$}
 \put(-225,75){\large$J/p$}
  \put(-325,-10){\large$d_{\alpha\beta}$}
    \put(-100,-10){\large$d_{\alpha\beta}$}
     \put(-250,125){\Large (a)}
  \put(-180,125){\Large (b)}
\caption{Protein synthesis rate $J/p$ as a function of the filament end-to-end distance $d_{\alpha\beta}$ for parameters  $\alpha_{\infty}/p = 0.4$, $\beta/p=1$  for $D_{\rm eff}=1$ [Panel(a)] and $D_{\rm eff}=0.1$ [Panel (b)].   Theoretical result Eq.~(\ref{eq:J3D})  for filaments in $\mathbb{R}^2$ ($d=2$, red solid lines) and $\mathbb{R}^3$ ($d=3$, black dashed lines) are compared with simulations results for filaments consisting of  $\ell=100$ monomers (markers).  The theoretical result Eq.~(\ref{eq:J3D}) applies for $d_{\alpha\beta}>2$ and $J=0.24$ for $d_{\alpha\beta} =0$.   Therefore, we have added dotted lines connecting  $J=0.24$  for $d_{\alpha\beta} =0$ with $J$ at $d_{\alpha\beta}=2 $.  The remaining parameters that specify the simulations can be found in Sec.~\ref{Sec:Simulation}.}\label{phase_diagram2}
\end{figure}
% -------------------------------------------------------------------------------------

% -------------------------------------------------------------------------------------
\begin{figure}
\includegraphics[width=0.4\textwidth]{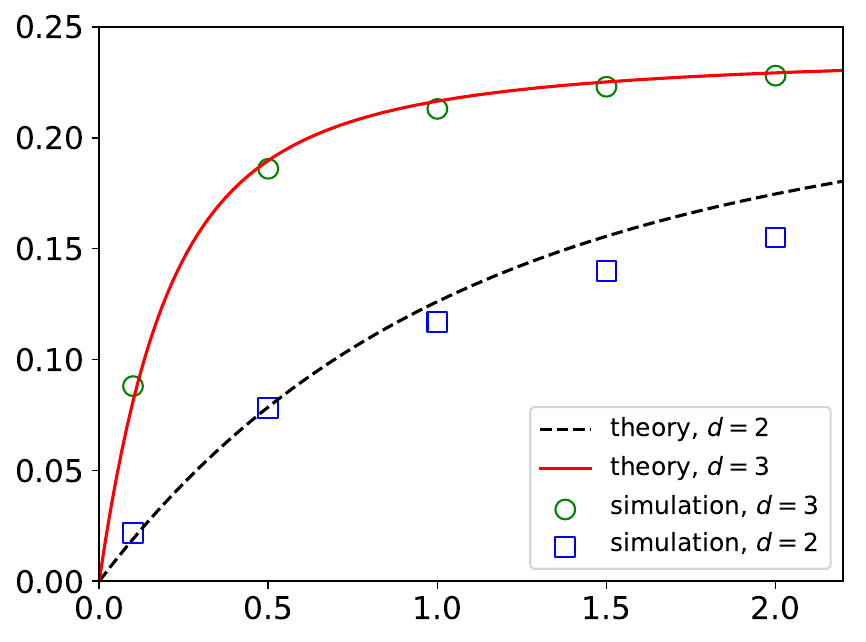}
 \put(-228,75){\large$J/p$}
  \put(-95,-10){\large$D_{\rm eff}$}
\caption{Protein synthesis rate $J/p$ as a function of the effective diffusion contant $D_{\rm eff}$  for filaments in  $\mathbb{R}^2$ ($d=2$) and $\mathbb{R}^3$  ($d=3$).   Analytical results from mean-field theory [solid lines depicting Eq.~(\ref{eq:J3D}) with $\mu_d$ as in Eqs.~(\ref{eq:mu2x}) or (\ref{eq:mu3x})] are compared with simulation results (circles).     The  parameters  used to compute the theoretical curves are 
$d_{\alpha\beta} = 20$, $\alpha_{\infty}/p = 0.4$, and $\beta/p>1/2$ (and therefore $\lim_{D_{\rm eff}\rightarrow \infty}J/p = 0.24$).   The remaining parameters that specify the simulations can be found in Sec.~\ref{Sec:Simulation}.   
}\label{currentDiffusion}
\end{figure}
% -------------------------------------------------------------------------------------

\subsection{Summary of the theoretical results} %--------------------------------------------

Using a mean-field approximation, we have derived the formula (\ref{eq:J3D}) for the current $J$  in the RTD model, which describes the   
protein translation rate for  one mRNA in a  diffusive reservoir that contains a large number of ribosomes.    The formula Eq.~(\ref{eq:J3D}) expresses the protein translation rate in terms of five parameters:   the elongation rate $p$; the ratio $\beta/p$ between the termination rate $\beta$ and $p$; the ratio $\alpha_{\infty}/p$ between the initiation rate  $\alpha_{\infty}$ in a  homogeneous reservoir, i.e. with an infinite diffusion constant $D$, and $p$; an effective diffusion constant $D_{\rm eff}$; and  a dimensionless parameter $\mu_d$ that quantifies the effect of the geometry of the setup on $J$.    For small values of the parameter $\zeta$, given by Eq.~(\ref{eq:zeta}), we obtained the simpler expression (\ref{noExclu}), which is independent of the exit rate $\beta$ and the hopping rate $p$. 

Interestingly,  all the geometrical details of the problem, such as the shape of the reservoir, the position of the filament in the reservoir, and the filament end-to-end distance, are captured by the parameter  $\mu_d$.   In order to understand the effect of dimensionality on $\mu_d$, we have determined $\mu_d$ for  the Euclidean spaces $\mathbb{R}^2$ and $\mathbb{R}^3$.    Surprisingly, the functional dependency of  $\mu_d$ on the end-to-end distance $d_{\alpha\beta}$ is qualitatively different  in two than in three dimensions.   In two dimensions, we find that $\mu_2$ diverges for large $d_{\alpha\beta}$, while in three dimensions $\mu_3$ converges to a finite value for large $d_{\alpha\beta}$.  This implies that  in two dimensions long mRNA filaments will have a vanishing protein translation rate, while in three dimensions the protein translation rate will be finite for long mRNA filaments.     In addition, in order to understand the effect of confinement on $\mu_d$, we have determined $\mu_d$ for a rectangle and a cuboid.  Explicit computations for the cuboid show that confinement effects disappear rapidly for  linear dimensions larger than $10r$, with $r$ the size of the reaction volume. 

So far, all results are based on mean-field theory.   In the next section, we validate  mean-field theory predictions   with simulations results for the RTD model.

% -------------------------------------------------------------------------------------
\section{Comparing mean-field theory with simulations}\label{Sec:Simulation}
% -------------------------------------------------------------------------------------

We have performed numerical simulations of the RTD to verify  the accuracy of the  mean field theory given by Eq.(\ref{eq:J}).   First, we we detail the specifics of the Monte Carlo simulations.   In a second subsection, we discuss the parameters used in the simulations.  In a final subsection, we compare predictions from mean-field theory with simulation results.

 \begin{figure}
\includegraphics[width=0.4\textwidth]{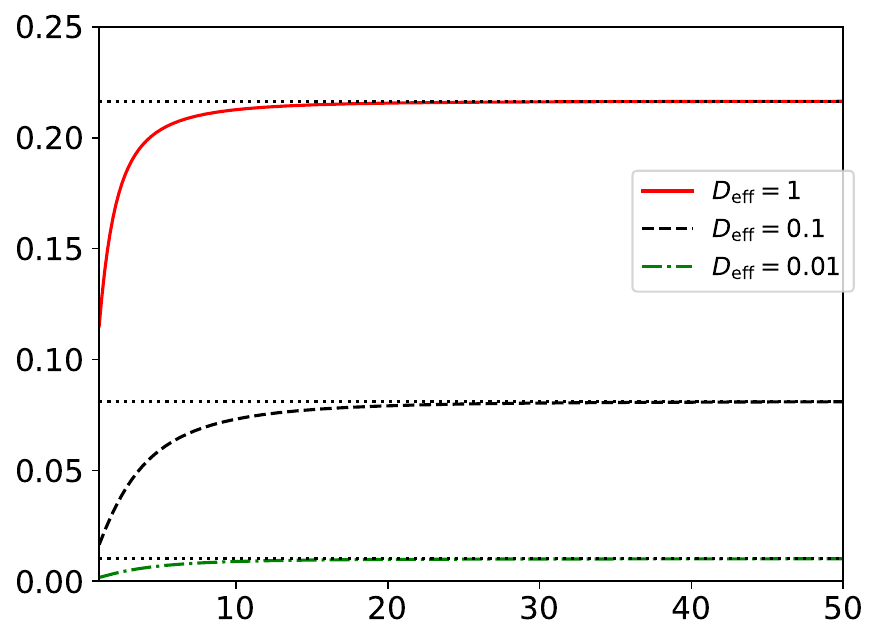}
 \put(-228,75){\large$J/p$}
  \put(-95,-10){\large$L_z/r$}
\caption{  Protein synthesis rate $J/p$ as a function of  $L_z$   for filaments in  a threedimensional cuboid with $L_x = \infty$ and $L_y=\infty$.   The filament is located in the $x,y$-plane and $L_z$ is thus an orthogonal direction.   Lines are analytical results from mean-field theory for a given value of $D_{\rm eff}$.    The  parameters  used to compute the theoretical curves are the same as in Fig.~\ref{currentDiffusion}, namely, 
$d_{\alpha\beta} = 20$, $\alpha_{\infty}/p = 0.4$, and $\beta/p>1/2$.      The dotted lines denote the asymptotics  for $L_z=\infty$ and are the same as in Fig.~\ref{currentDiffusion}. 
}\label{currentDist}
\end{figure}
 
% --------------------------------------------------------------------------------
\subsection{Monte Carlo simulations of the RTD} % ----------------------------------
% -------------------------------------------------------------------------------- 
Both components of the RTD, i.e., diffusion of particles  and the active transport on the filament, can be simulated independently, using a continuous-time Monte Carlo simulation on the TASEP~\cite{Newman,Walter15} and a Brownian motion in the reservoir.  However, in order to simulate the RTD model, we need to couple the dynamics of the two processes. 

In this subsection, we describe the algorithm used to simulate the dynamics of ribosomes in the reservoir, ribosomes on the filament, and how these two dynamics are coupled at the first and last site of the filament, where the ribosomes respectively enter on and exit from the filament.

First, we detail the simulations of the unbound ribosomes diffusing in the reservoir. We consider that unbound ribosomes do not interact with each other and their positions $\vec r$ evolve according to a Brownian equation of motion 
\begin{equation}
\frac{d\vec r}{dt}=\vec\xi(t)\,,
\label{Brownian_eq}
\end{equation}
where $\vec\xi$ is a white noise with
\begin{eqnarray}
 \langle\xi_{a}(t)\rangle&=&0\,,\\
 \langle\xi_a(t)\cdot\xi_b(t')\rangle&=&2D\delta(t-t')\delta_{a,b}\,,
\end{eqnarray}
where the indices $a$ and $b$ stand for the spatial coordinates of the ribosomes, i.e., $x$ and $y$ for a two-dimensional reservoir; and $x$, $y$ and $z$ for three-dimensional reservoir. 
We numerically integrate these equations  by discretizing time into intervals of length $\Delta t=t-t'$, namely, 
\begin{equation}
\frac{\vec r (t+\Delta t)-\vec r(t)}{\Delta t}=\vec\xi(t)\,.
\label{Brownian_eq2}
\end{equation}
The $\delta(t-t')$ in the amplitude of the white noise is replaced by $1/\Delta t$, leading to the following update rule for each spatial coordinate, 
\begin{equation}
r_a (t+\Delta t)=r_a(t) + \sqrt{2D\Delta t}\,\xi_a\,.
\label{Brownian_eq3}
\end{equation}
The reflecting boundary conditions are implemented as follows: if the update of a Brownian particle leads to a position outside of the box, the move is rejected.

%where $\eta_a$ is a random number uniformly distributed between 0 and 1.

Second, we detail the simulations of ribosomes bound to a filament located inside the reservoir. The filament contains $\ell$ sites and each site has the length $r$ of a ribosome. The filament has thus a total length $L=\ell r$. The dynamics on the TASEP is performed with a continuous-time Monte Carlo algorithm~\cite{Newman,Walter15}, sometimes called Gillespie algorithm~\cite{Gillespie}.  A configuration of ribosomes on the filament allows only for a finite number of moves given by the TASEP rules described above. For illustration, in the particular case of Fig.1, the first site is empty, thus a ribosome can enter at a rate $\alpha=\tilde\alpha N(t)$; three ribosomes  on the filament are free to move at a rate $p$ as there is no ribosome blocking passage in front of them; and finally a ribosome occupies the exit site of the filament, so it can leave at the rate $\beta$ the filament and return to the reservoir to resume a Brownian motion. It is useful to define the sum $S_r$ of the possible transition rates; in the case of Fig.1, $S_r=\alpha+\beta+3p$. A particular move is chosen with a probability linearly related to its rate and the filament is forced to perform this move. For instance, the probability $P_\beta$ to move the ribosome from the exit site to the reservoir is $P_\beta=\beta/S_r$.  The continuous-time Monte Carlo algorithm thus avoids rejection of ribosome moves, which saves us a considerable amount of computational time when densities of ribosomes are large.  The other advantage of a continuous-time Monte Carlo is that it runs in continuous and time.  The time between two consecutive moves is a random variable $\tau$ taken from  an exponential distribution with  mean value  like $S_r^{-1}$.  The explicit definition of $\tau$ is important as it allows us to couple  dynamics of ribosomes on the filament with the dynamics of ribosomes diffusing in the reservoir. Note that, intuitively, the sum of rates $S_r$, and thus the time $\tau$ spent by the filament during a move,  depend on the configuration of ribosomes. If $S_r$ is small (large), i.e., if a transition is unlikely (resp. likely) to happen, then the time evolution of the filament will be large (resp. small).

Third, we discuss how the dynamics in the reservoir is coupled to transport on the filament. First we draw a time $\tau$ from the continuous-time Monte Carlo algorithm, then we update the reservoir configuration over this time interval by integrating the Brownian equations for each particle in the reservoir over thle time $\tau$, and then we draw another time  $\tau$ and so on.  Hence, in this approach, we assume that in the time $\tau$ the reservoir does not change significantly. The internal dynamics of ribosome hopping is by definition not coupled to the reservoir as, in the RTD the ribosomes can neither attach nor detach in the bulk of the filament. The coupling between reservoir and filament takes place  at the first and last site of the filament. Therefore, it is sufficient to define the positions $\mathbf{r}_{\alpha}$ and $\mathbf{r}_{\beta}$ of the first and the last sites, respectively. Note that the end-to-end distance $d_{\alpha\beta}=|\mathbf{r}_{\beta}-\mathbf{r}_{\alpha}|$  can take any value between $0$ and $L$ depending on the conformation of the filament.   Among the possible moves accounted in the simulation is the attachment of a ribosome at the entrance: we define a spherical reaction volume $\mathcal{V_\alpha}=4/3\pi r^3$ of radius $r$ centered at the first site of the TASEP. If an unbound ribosome  is present in the reaction volume $\mathcal{V_\alpha}$ then it can attach at a rate $\tilde\alpha$ to the fiament, with $\tilde\alpha$ defined in Eq.(\ref{eq:alpha}). In the same way, a spherical volume $V_\beta$ of radius $r$ is centered at the exit site of the filament. If a ribosome exits the filament at a rate $\beta$, then it is released at a random position inside $V_\beta$, after which it resumes a Brownian motion in the reservoir. Note that we have used the same numerical technique  in Ref.~\cite{Cian} to coupled the TASEP-LK with Brownian particles inside a reservoir. 

%-----------------------------------------------------------------
\subsection{Parameters of the simulations} %----------------------
%-----------------------------------------------------------------

%We set $r=1$ and $p=1$.  This boils down to measuring distances in units of $r$ and measuring time intervals in units of $1/p$.
We describe in this paragraph the parameters chosen in the simulations.  

We first discuss the geometrical parameters of the simulation.  The filament has a  length $L=\ell\:r$ with $\ell=100$, which for a TASEP model is  large enough to keep finite size effects of the order of a few percents~\cite{Derrida92,Kolomeisky}.  In the simulations, the filament is located in the middle of the reservoir to ensure isotropy of the particle concentration and limiting boundary effects.  The reservoir is chosen  large enough with respect to $d_{\alpha,\beta}$ and $r$. In three dimensions, we choose the dimensions   orthogonal direction to the filament equal to 
$L_x=L_y=100\:r$, whereas the longitudinal direction parallel to $d_{\alpha,\beta}$ is taken to be larger, i.e., $L_z=200\:r$.   In two dimensions, we set $L_x=400\:r$ in the longitudinal direction and $L_y=200\:r$ in the orthogonal direction.   Note that the gradient of ribosomes in the reservoir induced by the transport on the filament is expected to be larger along the longitudinal direction to $d_{\alpha,\beta}$. This is why the longitudinal  dimension is chosen larger than the orthogonal directions.  With these reservoir dimensions, boundary effects are small as the system is large with respect to the gradient of particles.  Indeed, from Fig.~\ref{currentDist} we can conclude that for  linear dimensions larger than $20r$ the effects of confinement are negligable and the reservoir can be considered infinitely large.

We now discuss the remaining parameters of the system linked to the concentration of ribosomes, attachment rate at the entry site of the filament, and the diffusion coefficient of the Brownian motion.  In three dimensions, the system contains $10^5$ ribosomes, leading to a density of ribosomes $c_\infty=0.05\:r^{-3}$; whereas in two dimensions, we set the total number of ribosomes equal to $5\times10^5$, leading to a density $6.25\:r^{-2}$. In two and three dimensions, we used the same parameters. We chose $\tilde\alpha=0.4/c_\infty$ so that $\alpha_\infty=\tilde\alpha\,c_\infty=0.4$. We chose $D=0.1 \tilde\alpha r^2$ and $D=\tilde\alpha r^2$ in Fig.3(a) and (b), so that $D_{\rm eff}=D/(\tilde\alpha r^2)=0.1$ and $D_{\rm eff}=1$, respectively. Finally, we chose $\beta=p=1$.

%We locate the filament  in the middle of the reservoir (to ensure isotropy of possible small finite size effects), and its direction is longitudinal to the $z-$direction of the box.  {[\bf  IN: this only makes sense in three dimensions? ]}The plane $(x,y)$ thus constitutes the section plane of the TASEP. 
In this paragraph, we discuss how we choose the correlation and equilibration times in order  to increase the quality of the sampling during Monte Carlo simulations. The correlation time can be approximated by the time needed to replace all the ribosomes on the TASEP. During one MC iteration, the time spent during the update is $\tau\sim 1/S_r\sim 1/(\rho \ell p)$ where $\rho$ is the global density of ribosomes on the filament, i.e., $\rho=N_r/\ell$ where $N_r$ is the total number of ribosomes on the filament. Note that, in the last approximation of $\tau$, the sum of the rates $S_r$ is obtained assuming that it is dominated by the hopping rates in the bulk of the TASEP, which contains $\approx\rho\ell$ particles. The last ribosome that entered the TASEP will need at least to be chosen $\ell$ times amongst $\rho \ell$ possibilities of moves. Therefore the correlation time becomes $\tau_c\approx \rho \ell^2\tau=\ell/p$. As $p=1$ and $\ell=100$ in our simulations (both two and three dimensions), we use $\tau_c=100$. Starting with an empty initial configuration, we ensure the steady state by performing $100\tau_c = 10^4$ iterations described above (continuous time on the filament and integration of the Brownian motion in the reservoir). Subsequently, $2\times10^4$ samplings are performed in three dimensions and $10^5$ samplings in two dimensions, each spaced by $\tau_c=100$ iterations to decorrelate the configurations.  This leads to errors bars smaller than symbols.

\subsection{Results}
In Figs.~\ref{phase_diagram2} and \ref{currentDiffusion} we compare mean-field theory with results from  numerical simulations.    Theory and simulations are in very good correspondence, despite the fact that theory neglects correlations between particles, finite size effects on the filament due to boundary layers,   and confinement  effects due to the finite  volume of the reservoir.  The very good correspondence between numerical experiments and theory shows that the expression for the current $J$ given by ~Eqs.~(\ref{eq:alphaFinal}-\ref{eq:J3D}) describes well the effect of     finite mobility on the protein synthesis rate $J$.  
In Figures~\ref{phase_diagram2} and \ref{currentDiffusion}, we observe that simulations overestimate the protein synthesis rate at small values of $D_{\rm eff}$ and underestimate the current at intermediate values of  $D_{\rm eff}$.  We expect that these deviations are due to correlation between particles inside the reaction volume and on the filament.

% ----------------------------------------------------------------------------------------------
\section{Biological relevance of diffusion in ribosomal recycling}     \label{Sec:param}
%-----------------------------------------------------------------------------------------------

To determine the biological relevance of finite mobility for ribosomal recycling, we use experimentally measured values for the parameters that appear in the theoretical expression for the protein synthesis rate derived in Sec.~\ref{Sec:theory}. We focus on two organisms for which the required microscopic parameters have been measured experimentally, namely, the bacterium  {\it Escherichia coli} and the budding yeast {\it Saccharomyces cerevisiae}. Moreover, we focus on the three-dimensional case corresponding to cytoplasmic translation.      

\begin{table}[h!]
  \begin{center}
    \caption{Impact of finite mobilities on ribosomal recycling in two  organisms.}
    \label{tab:table1}
    \begin{tabular}{c|c} % <-- Alignments: 1st column left, 2nd middle and 3rd right, with vertical lines in between
      \textbf{{\it E. coli} } & \textbf{{\it S. cerevisiae}} \\
      \hline
$\mu_3/D_{\rm eff} < 0.06$  & $\mu_3/D_{\rm eff} < 0.007$ 
    \end{tabular}
  \end{center}
\end{table}

 Since for physiological parameters the initiation of translation is the rate limiting step, we use the expression for the protein synthesis rate given by Eq.~(\ref{noExclu}).  Eq.~(\ref{noExclu}) implies that if
 \begin{eqnarray}
 \mu_d \ll D_{\rm eff} 
 \end{eqnarray}
 then diffusion has no meaningful influence on the protein synthesis rate. On the other hand, when 
 \begin{eqnarray}
 \mu_d \gg D_{\rm eff} 
 \end{eqnarray} 
then the influence of finite diffusion on protein synthesis rate is sizeable. Hence, in what follows we estimate the parameters  $\mu_d$ and  $D_{\rm eff}$.        

% ------------------------------------------------------------------------------------------------
\subsection{Estimate of  $\mu_3$} %---------------------------------------------------------------
%-------------------------------------------------------------------------------------------------
First, we estimate  the geometric parameter $\mu_3$ corresponding to cytoplasmic translation. Formula (\ref{eq:mu3x}), implies   for a three dimensional and infinitely large reservoir that
   \begin{eqnarray} 
\mu_3 \leq \frac{2}{5},
\end{eqnarray}
where  the equality is achieved in the limit $d_{\alpha\beta}\rightarrow\infty$.

%In two dimensions, using that the ribosome has a radius $r$ larger than $10 {\rm nm}$ and that for a typical mRNA with nucleotides $d_{\alpha\beta}\approx 300 {\rm nm}$ \cite{Phil}, we obtain 
%\begin{eqnarray}
%\mu_2 \leq  \frac{\log \frac{300}{10}+1}{2} \approx 2.2.
%\end{eqnarray}

% ------------------------------------------------------------------------------------------------
\subsection{Estimate for $D_{\rm eff}$ in Escherichia coli} %-------------------------------------------------------------------
%------------------------------------------------------------------------------------------------
In order to  estimate  $D_{\rm eff}$, it is useful to rewrite the expression Eq.~(\ref{eq:DEff}) in terms of  $\langle \alpha \rangle $, which gives 
\begin{eqnarray}
D_{\rm eff} = \frac{D \langle N_r\rangle }{ \langle  \alpha \rangle r^2} \label{eq:Deffx}
\end{eqnarray}
where $\langle  N_r\rangle$ is the number of ribosomes in the reaction volume.      According to Eq.~(\ref{eq:J3dFinal}), $\langle N_r\rangle$ is lower bounded by 
\begin{equation}
\langle N_r\rangle > \frac{4\pi}{3}r^3c_{\rm u}  - \frac{Jr^2}{2D} \label{eq:Nr}
\end{equation}
where $c_{\rm u}$ denotes the concentration of unbound ribosomes. The second term in Eq.~\ref{eq:Nr} is a correction due to depletion of ribosomes around the entry site.    Substituting Eq.~(\ref{eq:Nr}) in Eq.~(\ref{eq:Deffx}) and using $J = \langle \alpha\rangle$,  we obtain. 
\begin{eqnarray}
D_{\rm eff} > \frac{4\pi}{3}\frac{D c_{\rm u} r}{ \langle  \alpha \rangle} - \frac{1}{2}. \label{eq:Deffxx}
\end{eqnarray}
The quantity $\langle \alpha \rangle $ is hard to estimate since it can vary in several orders of magnitude from one mRNA transcript to another, see for instance Ref.~\cite{Shah}. However, since initiation is the rate limiting step,  it holds that 
\begin{eqnarray}
\langle\alpha  \rangle < \frac{p}{2},  \label{eq:boundingInitiation}
\end{eqnarray}
with the elongation rate $p$ being fairly independent of the mRNA transcript and the biological organism under study.   Combining Eqs.~(\ref{eq:Deffx}) and (\ref{eq:boundingInitiation}), we obtain the lower bound
\begin{eqnarray}
D_{\rm eff}  >  \frac{8\pi}{3}\frac{D c_{\rm u} r}{ p} - \frac{1}{2}. \label{eq:DeffBound}
\end{eqnarray}

We are left to estimate the parameters $D$, $c_{\rm u}$, $r$ and $p$.  We first consider the case of the bacteria {\it Escherichia coli}.   

Empirical values for the diffusion of ribosomes in {\it E. coli} show that $D\approx 0.04\mu m^2/s$, see Table 4-1 in Ref.~\cite{Phil}.    However, the diffusion coefficient of the subunits of unbound ribosomes (i.e., those not bound to mRNA), is one order of magnitude larger and given by $D\approx 0.2\mu m^2/s$, as shown in Ref.~\cite{Single}.  

For the radius of the reaction volume $r$, we  use that the reaction volume cannot be smaller than the radius of a ribosome (or one of its subunits), and thus $r>10 {\rm nm}$, see Figure 1-40 in Ref.~\cite{Phil}.  
 %Because of potential longer range interactions, we may set $r < 50 {\rm nm}$ such that it is a sizeable fraction of the size of the mRNA, a matter of being very generous with our estimates.   

 For {\it E. coli}, the elongation rate $p$ has been measured in several experiments, see Refs.~\cite{Young, Bilgin, Proshkin}, leading to a value $p$ of about about $10-20$ codons per second.    Since a ribosome occupies three codons, we take for $p\approx 7 s^{-1}$.
 
Lastly, we need an estimate for the concentration
\begin{eqnarray}
c_{\rm u} = \frac{N_{\rm u}}{V}.  \label{cu}
\end{eqnarray} 
The volume of {\it E. coli} is $V \approx 1 \mu m^3$ and its total number of ribosomes is about $N_{\rm tot} = 20000$~\cite{Phil}.  The fraction of unbound (or free) ribosomes is  about 15\% \cite{Single,Forchhammer} of the total value, leading to 
\begin{eqnarray}
c_{\rm u} \approx  2\times 0.15 \times 10^4\mu m^{-3} \approx 3 \times 10^3\mu m^{-3} . \label{EcoliCu}
\end{eqnarray}

Combining all parameter values into the right hand side of the bound Eq.~(\ref{eq:DeffBound}) for $D_{\rm eff}$, we obtain that 
\begin{eqnarray}
D_{\rm eff} > \frac{8\pi}{3}\frac{0.2 \times 10\times 3 \times 10^3  }{ 7 }    \frac{{\rm nm}}{\mu m} \approx 6.8 ,
\end{eqnarray} 
and therefore
\begin{eqnarray}
\frac{\mu_3}{D_{\rm eff}} < 0.06 .
\end{eqnarray}

We can conclude that diffusion has no sizeable effect on  protein synthesis rates. This is in particular true since we have been very generous with all the biological parameters. For example, taking $\langle \alpha\rangle < p/20$ instead of $p/2$, as in Ref.~\cite{CiandriniSTan}, would provide an even smaller upper bound $\frac{\mu_3}{D_{\rm eff}} < 0.006$.

%------------------------------------------------------------------------------------------------
\subsection{Estimate for $D_{\rm eff}$ in Saccharomyces cerevisiae} %----------------------------------------------------------
%------------------------------------------------------------------------------------------------
As a second example, we consider the case of  budding yeast.  We use again Eq.~(\ref{eq:DeffBound}) to bound $D_{\rm eff}$.   All empirical values are known for this organism, see for instance table S1 in Ref.~\cite{Shah}. 

Empirical values for the diffusion coefficient of the 60S subunit of ribosomes in the dense nucleoplasm of budding yeast show that $D\approx 0.3(\mu m)^2/s$~\cite{Politz}. We may expect that ribosomes diffuse faster in the cytoplasm, where translation takes place. 

For the radius of the reaction volume $r$, we  use again the reaction volume cannot be smaller than the radius of the ribosome, and thus  $r > 10 {\rm nm}$.  

The elongation rate of ribosomes in budding yeast has been measured to be $p\sim 10 $ codons per second and therefore $p\approx 3 s^{-1}$ since a ribosome occupies three codons \cite{Shah, Arava}. 

Finally, we come to the estimate of $c_{\rm u}$, given by Eq.~(\ref{cu}). The volume of a budding yeast cell is about $V \approx 42\mu m^{3}$ \cite{Shah, Siwiak} and the number of ribosomes is $2\times 10^5$ \cite{Shah, Warner, Haar}. Using again that a fraction $15\%$ of ribosomes are unbound, see Figure 3 in \cite{Shah}, we obtain 
 \begin{eqnarray}
 c_{\rm u} \approx  \frac{2 \times 0.15 \times 10^5}{42}\mu m^{-3} \approx  7  \times 10^3\mu m^{-3} ,
 \end{eqnarray}
 which is in fact close to the concentration of unbound ribosomes in {\it E.~coli}, see Eq.~(\ref{EcoliCu}). 
 
 Combining all parameters in the bound given by Eq.~(\ref{eq:DeffBound}), we obtain that 
\begin{eqnarray}
D_{\rm eff} > \frac{8\pi}{3}\frac{0.3 \times 10\times 7 \times 10^3  }{ 3 }    \frac{{\rm nm}}{\mu m}  - 1/2 \approx 59
\end{eqnarray}
and 
 \begin{eqnarray}
\frac{\mu_3}{D_{\rm eff}} < 0.007 .  \label{eq:SC}
\end{eqnarray}  

We should again bear in mind that the bound in Eq.~(\ref{eq:SC}) is  a generous upper bound  based on the bound on the initiation rates given by Eq.~(\ref{eq:boundingInitiation}), and  it is thus likely a loose bound and a significant overestimate for $\mu_3/D_{\rm eff}$.

\subsection{Protein synthesis rates  for   {\it E. coli}}
\begin{figure}
\includegraphics[width=0.4\textwidth]{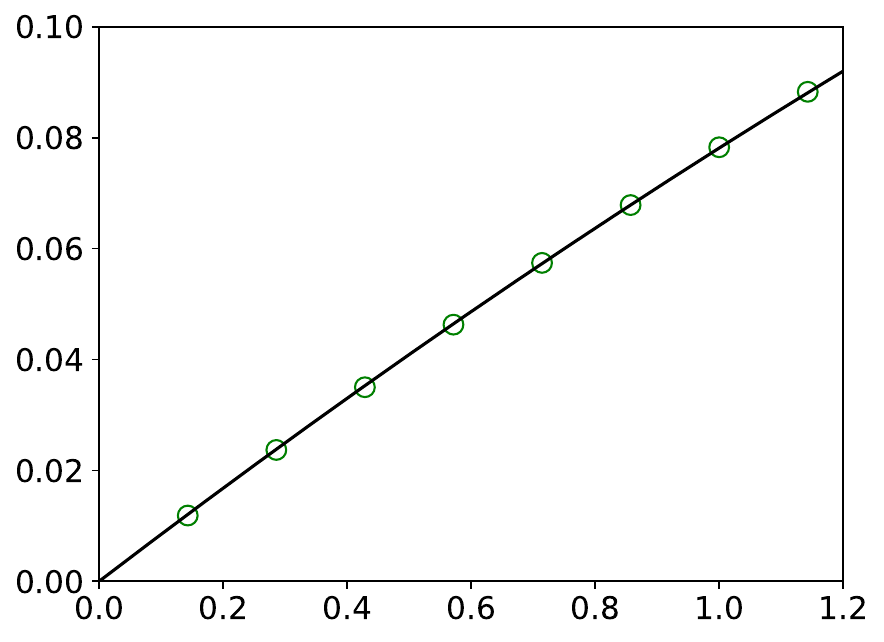}
 \put(-228,75){\large$J/p$}
  \put(-95,-10){\large$\tilde\alpha/p$}
\caption{We plot $J/p$ versus $\tilde\alpha/p$ for biological parameters for {\it E. coli}: $D=2\mu m^2s^{-1}$, $L_x=L_y=0.7\mu m$ and $L_z=2\mu m$ so that $V\approx 1\mu m^3$, $p=\beta=7s^{-1}$, the radius of the reaction volume is $r=10nm$, the length of the mRNA is $\ell=$300 sites corresponding to codons, and $d_{\alpha,\beta}=300nm$ and we took 20.000 ribosomes in the reservoir. Results from simulations are given by symbols and compared to $J_\infty=\alpha_\infty/p(1-\alpha_\infty/p)$ corresponding to infinite diffusion constant.}
\label{biology}
\end{figure}

We end this section by presenting Figure~\ref{biology} that shows simulation results for the protein synthesis rate $J/p$ in the parameter regime that is relevant for mRNA translation in {\it E. coli}. We compare these results from simulations with $J_\infty/p$, the standard TASEP result for $D_{\rm eff} = \infty$. We see that both are indistinguishable and confirms that finite diffusion is not a limiting factor in mRNA translation.    

%------------------------------------------------------------------------------------------------------
\section{Discussion}   \label{Sec:disc} % -------------------------------------------------------------
%------------------------------------------------------------------------------------------------------

We have made a study of a  totally asymmetric simple exclusion process  immersed in a diffusive reservoir~\cite{Blythe, Chou2011}, which we have called the RTD model. The RTD is a model for translation based on directed transport of ribosomes along mRNA and recycling of ribosomes through diffusion in the cytoplasm. We have used this model to determine whether under physiological conditions  diffusion is a limiting factor for ribosome recycling.

We have derived an analytical expression for the current $J$ at which mRNA is translated into proteins, which is corroborated by  numerical simulation results.  These results show that finite diffusion leads to a reduction in the translation rate $J$   because the concentration of ribosomes at the mRNA initiation site is depleted.  In addition, we find that the ratio between a geometric parameter $\mu_d$ and an effective diffusion coefficient $D_{\rm eff}$ determines whether diffusion has an impact on the protein synthesis rate: if $\mu_d \ll D_{\rm eff}$, then the concentration of ribosomes at the 5' end of the mRNA is not affected by finite diffusion; on the other hand, if $\mu_d\gg D_{\rm eff}$ then  depletion of ribosomes at the mRNA initiation site is significant.   

Using a broad range of physical parameters, we find that it is unlikely that finite diffusion  is a limiting factor under physiological conditions in ribosome recycling. Indeed, in Table~\ref{tab:table1}, we present generous upper bounds for the parameter $\mu_d/D_{\rm eff}$ for two organisms, namely, the bacterium {\it E. coli} and the yeast {\it S. cerevisiae}. In both cases, we obtain that $\mu_d/D_{\rm eff}$ is substantially smaller than~$1$.  

The outcome of our analysis, namely that the finite mobility of ribosomes does not play a role in translation control, is not a complete surprise given  that ribosomes diffuse at large enough rates. For example, it takes $0.1 \: {\rm s}$ for a protein to diffuse across an {\it E. coli} cell and $10 \: {\rm s}$ for a protein to diffuse across a yeast cell~\cite{Alon}, while the time to translate a protein is about $2 \: {\rm min}$~\cite{Alon}. Hence, as much as concerns the translation of mRNA into proteins, the diffusion rate of ribosomes can be considered very large and therefore of negligeable effect on  the whole translation process.     Also, since ribosomes biogenesis is one of the most resource expensive processes for  cells~\cite{Woolford, Warner}, it is reasonable to assume that the molecular conditions are optimised by evolutionary constraints in order to render translation efficient, which in the present context implies that translation is not limited by ribosome mobilities.

From a biological point of view, these results imply that the purpose of  mRNA circularisation~\cite{LodishEight, Wells} is not the optimization of ribosome recycling by reducing the limiting factor of  diffusion in the cytoplasm. Instead, the circularisation of mRNA may regulate the efficiency of translation initiation by altering the binding strength of initiation factors to the mRNA~\cite{Wells, Vicens}.
%Also, the length-dependence of ribosomal densities on mRNA~\cite{Fer2017} is likely not due to the finite diffusion of ribosomes in the cytoplasm. An alternative explanation may be  due to  ribosome drop-off rates, see Ref.~\cite{Bonnin2017} \red{[this is addressed by Fernandes et al I think ?]}, or due to an increase in mRNA translation efficiency after circularisation~\cite{Guo} \red{[the last arguments is not obvious to me]}. 
Hence, we come to a different conclusion than Ref.~\cite{Fer2017}, which argues that three-dimensional diffusion of ribosomes in the cytoplasm plays an important role for mRNA translation control. Note that the question of the effect of the finite mobility of ribosomes on the current on mRNA remains open in two dimensions, as the diffusion coefficient of ribosomes constrained to a two-dimensional diffusion on the endoplasmic reticulum is not known to our knowledge.

Although finite diffusion is not   rate limiting for ribosome recycling under physiological conditions, the  RTD model may be relevant to explain the reduction in protein production when cells are in a dormant state. The mobility of cytoplasmic particles in dormant yeast cells is much lower than their mobility in yeast cells under normal conditions~\cite{Joyner, Munder}. The reduction in mobility of cytoplasmic particles is due to a transition between a fluid-like to a solid-like phase of the cytoplasm, which is triggered by the  acidification of the cytosol~\cite{Munder}. The formula $J \sim D$ indicates that the protein synthesis rate  scales proportional to the particle mobility at low values of $D$. 

The RTD model is also interesting as a model for the  coupling between active transport and passive diffusion. Remarkably, the rate $J$  admits a universal form that depends on five parameters only: the elongation rate $p$, the ratio $\beta/p$ between the rate $\beta$ of termination and $p$, the ratio $\alpha_{\infty}/p$ between the initiation rate  $\alpha_{\infty}$ for a homogeneous reservoir (i.e., the limit of an infinitively fast diffusion) and $p$, an effective diffusion constant $D_{\rm eff}$, and  a dimensionless parameter $\mu_d$ that quantifies the effect of the geometry of the reservoir and the filament on the current $J$.  
 We have also found an interesting qualitative distinction between finite diffusion in two and three dimensions. In two dimensions, it holds that the current $J$ vanishes in the large distance  limit between the filament end points, while in  three dimensions this limit gives a finite current $J$. However, the decay towards zero of $J$ in two dimensions, which may be relevant for the endoplasmic reticulum translation, is logarithmically slow. 
 
We end the paper by discussing  the assumptions made by the RTD model and interesting future extensions of the present paper.  First, we have ignored the fact that ribosomes disassemble into two subunits in the cytoplasm~\cite{LodishEight}. Hence, in principle we should consider a reservoir with two types of particles. However, if the mRNA binding rate one of these subunits is rate limiting, then the predictions of our model would remain valid. Interestingly, experimental data indicates that in prokaryotes the binding of the 40S ribosomal subunit is thought to be the rate-limiting step of initiation~\cite{Hershey}.  Second, we have assumed that mRNA has zero mobility and we have also assumed that the end-points of the mRNA are immobile.  Nevertheless, including diffusion of the mRNA in the model  would not alter the main conclusions of this paper, since it would only reduce the effects of  finite diffusion  on the protein synthesis rate. Third, it is known that cytoplasmic particles  diffuse anomalously within living cells~\cite{Jeon, Tejedor, Felix} and therefore a model based on fractional Brownian motion is more appropriate~\cite{Felix}. However, the exponent of the anomalous diffusion is close to $1$ ($0.88$ for nanosilica particles of various sizes in yeast cells~\cite{Munder}), and therefore we expect it not to have a major impact on short length scales. It would nevertheless be interesting to analyse the dependence of $J$ on $d_{\alpha\beta}$ in this case.  

\acknowledgements

This work was supported in part by a “Mod\'elisation pour le Vivant” CNRS Grant CoilChrom (2019--2020), and the LabEx NUMEV (ANR-10-LABX-0020) within the I-SITE MUSE of Montpellier University [No. AAP 2013-2-005, and Flagship Project Gene Expression Modeling (2017--2020)].

\appendix 

\section{Concentration of ribosomes in the box}\label{appendixA}
We solve Eqs.~(\ref{eq:Diffusion})-(\ref{eq:PI}) in various geometries when $|\mathbf{r}_{\beta} - \mathbf{r}_{\alpha}| > 2r$.

In $\mathbb{R}^2$, we obtain that
 \begin{eqnarray}
c(\bf r)=  \left\{\begin{array}{ccc}  c_{\infty}+\frac{Jr^2}{4D\mathcal{V}}-\frac{J}{4D\mathcal{V}}|{\bf r}-{\bf r}_\beta|^2+\frac{Jr^2}{2D\mathcal{V}}\ln\left(\frac{|\bf r-\bf r_\alpha|}{r}\right),\quad |{\bf r}-{\bf r}_\beta|<r,\label{c_exit2d}\\
c_{\infty}-\frac{Jr^2}{4D\mathcal{V}}+\frac{J}{4D\mathcal{V}}{|{\bf r}-{\bf r}_\alpha|^2}-\frac{Jr^2}{2D\mathcal{V}}\ln\left(\frac{|{\bf r}-{\bf r}_\beta|}{r}\right),\quad |{\bf r}-{\bf r}_\alpha|<r,\label{c_entrance2d}\\
c_{\infty}-\frac{Jr^2}{2D\mathcal{V}}\ln\left(\frac{|{\bf r}-{\bf r}_\beta|}{|{\bf r}-{\bf r}_\alpha|}\right),\quad |{\bf r}-{\bf r}_\alpha|>r, \ {\rm and } \  |{\bf r}-{\bf r}_\beta|>r \label{c_volume2d} ,\end{array} \right. %\label{eq:J}
\end{eqnarray} 
while in $\mathbb{R}^3$ it holds that
 \begin{eqnarray}
c(\bf r) =  \left\{\begin{array}{ccc}  c_\infty+\frac{Jr^2}{2D\mathcal{V}}-\frac{J}{6D\mathcal{V}}|{\bf r}-{\bf r}_\beta|^2-\frac{J r^3}{3D\mathcal{V}}\frac{1}{|\bf r-\bf r_\alpha|},\quad  |{\bf r}-{\bf r}_\beta|<r ,\label{c_exit3d}\\
c_\infty-\frac{Jr^2}{2D\mathcal{V}}+\frac{J}{6D\mathcal{V}}|{\bf r}-{\bf r}_\alpha|^2+\frac{Jr^3}{3D\mathcal{V}}\frac{1}{|{\bf r}-{\bf r}_\beta|},\quad |{\bf r}-{\bf r}_\alpha|<r ,\label{c_entrance3d}\\
c_\infty+\frac{Jr^3}{3D\mathcal{V}}\left(\frac{1}{|{\bf r}-{\bf r}_\beta|}-\frac{1}{|{\bf r}-{\bf r}_\alpha|}\right),\quad  |{\bf r}-{\bf r}_\alpha|>r, \ {\rm and } \  |{\bf r}-{\bf r}_\beta|>r \label{c_volume3d} .\end{array} \right. \label{eq:J3dFinal}
\end{eqnarray}

For a rectangular box of dimensions $L_x\times L_y$, we obtain 
 \begin{eqnarray}
c(\bf r)=  \left\{\begin{array}{ccc}  c_{\infty}+\frac{Jr^2}{4D\mathcal{V}}-\frac{J}{4D\mathcal{V}}|{\bf r}-{\bf r}_\beta|^2+\frac{Jr^2}{2D\mathcal{V}}\ln\left(\frac{|\bf r-\bf r_\alpha|}{r}\right) +  c_{\mathcal{I}}({\bf r}) ,\quad |{\bf r}-{\bf r}_\beta|<r,\label{c_exit2dbis}\\
c_{\infty}-\frac{Jr^2}{4D\mathcal{V}}+\frac{J}{4D\mathcal{V}}{|{\bf r}-{\bf r}_\alpha|^2}-\frac{Jr^2}{2D\mathcal{V}}\ln\left(\frac{|{\bf r}-{\bf r}_\beta|}{r}\right) +  c_{\mathcal{I}}({\bf r}) ,\quad |{\bf r}-{\bf r}_\alpha|<r,\label{c_entrance2dbis}\\
c_{\infty}-\frac{Jr^2}{2D\mathcal{V}}\ln\left(\frac{|{\bf r}-{\bf r}_\beta|}{|{\bf r}-{\bf r}_\alpha|}\right) +  c_{\mathcal{I}}({\bf r}) ,\quad |{\bf r}-{\bf r}_\alpha|>r, \ {\rm and } \  |{\bf r}-{\bf r}_\beta|>r \label{c_volume2dbis} . \end{array} \right. %\label{eq:J}
\end{eqnarray} 
where 
 \begin{eqnarray}
 c_{\mathcal{I}}({\bf r}) = \frac{Jr^2}{2D\mathcal{V}} \sum_{j\in \mathcal{N}_{\alpha}}\ln(|\mathbf{r}-\mathbf{r}^{(j)}_{\alpha}|) - \frac{Jr^2}{2D\mathcal{V}} \sum_{j\in \mathcal{N}_{\beta}}\ln(|\mathbf{r}-\mathbf{r}^{(j)}_{\beta}|) ,
\end{eqnarray} 
and where the  $\mathbf{r}^{(j)}_{\alpha}$ denote the coordinates of the images of the filament initiation site located at $\mathbf{r}_{\alpha}$, and where the $\mathbf{r}^{(j)}_{\beta}$ denote the coordinates of the images of the filament termination site $\mathbf{r}_{\beta}$, as illustrated in  Figure~\ref{fig:image}.

Analogously, for a cuboid of dimensions $L_x\times L_y\times L_z$, we obtain  
 \begin{eqnarray}
c(\bf r) =  \left\{\begin{array}{ccc}  c_\infty+\frac{Jr^2}{2D\mathcal{V}}-\frac{J}{6D\mathcal{V}}|{\bf r}-{\bf r}_\beta|^2-\frac{J r^3}{3D\mathcal{V}}\frac{1}{|\bf r-\bf r_\alpha|} +  c_{\mathcal{I}}({\bf r}),\quad  |{\bf r}-{\bf r}_\beta|<r ,\label{c_exit3dbis}\\
c_\infty-\frac{Jr^2}{2D\mathcal{V}}+\frac{J}{6D\mathcal{V}}|{\bf r}-{\bf r}_\alpha|^2+\frac{Jr^3}{3D\mathcal{V}}\frac{1}{|{\bf r}-{\bf r}_\beta|}+  c_{\mathcal{I}}({\bf r}),\quad |{\bf r}-{\bf r}_\alpha|<r ,\label{c_entrance3dbis}\\
c_\infty+\frac{Jr^3}{3D\mathcal{V}}\left(\frac{1}{|{\bf r}-{\bf r}_\beta|}-\frac{1}{|{\bf r}-{\bf r}_\alpha|}\right)+  c_{\mathcal{I}}({\bf r}),\quad  |{\bf r}-{\bf r}_\alpha|>r, \ {\rm and } \  |{\bf r}-{\bf r}_\beta|>r \label{c_volume3dbis} , \end{array} \right. %\label{eq:J}
\end{eqnarray}
where
 \begin{eqnarray}
 c_{\mathcal{I}}({\bf r}) =  -\frac{J r^3}{3D\mathcal{V}} \sum_{j\in \mathcal{N}_{\alpha}} \frac{1}{|{\bf r}-{\bf r}^{(j)}_\alpha|} + \frac{J r^3}{3D\mathcal{V}} \sum_{j\in \mathcal{N}_{\beta}} \frac{1}{|{\bf r}-{\bf r}^{(j)}_\beta|} .
\end{eqnarray}

 \end{document}